\documentclass[pdflatex,sn-mathphys-num]{sn-jnl}

\usepackage{graphicx}%
\usepackage{multirow}%
\usepackage{amsmath,amssymb,amsfonts}%
\usepackage{amsthm}%
\usepackage{mathrsfs}%
\usepackage[title]{appendix}%
\usepackage{xcolor}%
\usepackage{textcomp}%
\usepackage{manyfoot}%
\usepackage{booktabs}%
\usepackage{algorithm}%
\usepackage{algorithmicx}%
\usepackage{algpseudocode}%
\usepackage{listings}%

\theoremstyle{thmstyleone}%
\theoremstyle{thmstyletwo}%

\theoremstyle{thmstylethree}%

\begin{document}

\title[Hadronic contributions to $a_\mu$ within Resonance Chiral Theory]{Hadronic contributions to $a_\mu$ within Resonance Chiral Theory}


\author[1]{\fnm{Emilio J.} \sur{Estrada}}\email{emilio.estrada@cinvestav.mx}

\author[
1]{\fnm{Alejandro} \sur{Miranda}}\email{jmiranda@ifae.es}

\author*[1
]{\fnm{Pablo} \sur{Roig}}\email{pablo.roig@cinvestav.mx}

\affil*[1]{\orgdiv{Departmento de Física}, \orgname{Centro de Investigación y de Estudios Avanzados del Instituto Politécnico Nacional}, \orgaddress{\street{Apartado Postal 14-740}, 
\postcode{07360}, \state{Mexico City}, \country{Mexico}}}




\abstract{We review the recent progress achieved, using Resonance Chiral Theory, in the hadronic contributions to the muon anomalous magnetic moment. These include the hadronic vacuum polarization, either using $e^+e^-$ or $\tau$ decays into hadron final states as input; and the hadronic light-by-light part, where in addition to previous results on the lightest pseudoscalar and tensor-poles contributions, we first present the evaluation of the pseudoscalar box using this formalism. We also discuss the scalar, axial-pole and other subleading pieces. The results obtained are consistent with the White Paper 2 values, with comparable precision.}

\keywords{Muon anomalous magnetic moment, Resonances, Chiral symmetry, $1/N_C$ expansions, Short-distance QCD constraints.}



\maketitle

\section{Introduction}\label{sec_Intro}

The electron anomalous magnetic moment, $a_e=(g_e-2)/2$ was key in understanding better quantum corrections within field theory, as it is a purely quantum probe of the deviations, in the dipole coupling with a weak magnetic field, from the tree-level result stemming from the Dirac equation, $g=2$.

After early experimental effort and Schwinger's famous computation of $a_e=\alpha/(2\pi)+\mathcal{O}(\alpha^2)$, it became also possible to test this fundamental property for muons, first with the CERN experiments, reaching an accuracy of $8.5\times10^{-9}$~\cite{CERN-Mainz-Daresbury:1978ccd}, then at BNL~\cite{Muong-2:2006rrc}, improving up to an error of $6.3\times10^{-10}$, and finally with the FNAL experiment~\cite{Muong-2:2021ojo, Muong-2:2023cdq, Muong-2:2025xyk}, reducing further the uncertainty up to $1.5\times10^{-10}$. As a result of this Herculean effort, the current experimental average reads
\begin{equation}
a_{\mu}^{\rm exp} = 116592072(15)\times 10^{-11}\,.
\end{equation}

This astonishing precision has not yet been matched by the theory prediction within the Standard Model (SM). Its current value, as published in the second White Paper (WP) of the Muon g-2 Theory Initiative~\cite{Aliberti:2025beg}~\footnote{This one is based on Refs.~\cite{RBC:2018dos,Giusti:2019xct,Borsanyi:2020mff,Lehner:2020crt,Wang:2022lkq,Aubin:2022hgm,Ce:2022kxy,ExtendedTwistedMass:2022jpw,RBC:2023pvn,Kuberski:2024bcj,Boccaletti:2024guq,Spiegel:2024dec,RBC:2024fic,Djukanovic:2024cmq,ExtendedTwistedMass:2024nyi,MILC:2024ryz,Bazavov:2024eou,Keshavarzi:2019abf,DiLuzio:2024sps,Kurz:2014wya,Colangelo:2015ama,Masjuan:2017tvw,Colangelo:2017fiz,Hoferichter:2018kwz,Eichmann:2019tjk,Bijnens:2019ghy,Leutgeb:2019gbz,Cappiello:2019hwh,Masjuan:2020jsf,Bijnens:2020xnl,Bijnens:2021jqo,Danilkin:2021icn,Stamen:2022uqh,Leutgeb:2022lqw,Hoferichter:2023tgp,Hoferichter:2024fsj,Estrada:2024cfy,Ludtke:2024ase,Deineka:2024mzt,Eichmann:2024glq,Bijnens:2024jgh,Hoferichter:2024bae,Holz:2024diw,Cappiello:2025fyf,Colangelo:2014qya,Blum:2019ugy,Chao:2021tvp,Chao:2022xzg,Blum:2023vlm,Fodor:2024jyn,Aoyama:2012wk,Volkov:2019phy,Volkov:2024yzc,Aoyama:2024aly,Parker:2018vye,Morel:2020dww,Fan:2022eto,Czarnecki:2002nt,Gnendiger:2013pva, Hoferichter:2025yih}.},

\begin{equation}
a_{\mu}^{\rm SM} = 116 592 033(62)\times 10^{-11}\,,
\end{equation}
has an uncertainty entirely of hadronic origin and is completely dominated by that of the hadronic vacuum polarization piece, (HVP), of $61\times 10^{-11}$. The error of the Hadronic Light-by-light (HLBL) part has basically halved from the first WP~\cite{Aoyama:2020ynm}, up to the current $9.6\times 10^{-11}$ uncertainty~\cite{Aliberti:2025beg}. The HVP and HLbL contributions to the muon g-2, of order $\mathcal{O}(\alpha^{2,3})$, respectively, are illustrated in Figure~\ref{fig_HadronicContributions}.

\begin{figure}[h!]\label{fig_HadronicContributions}
\centering
\includegraphics[width=8cm]{HadConts}
\caption{Leading order (LO) hadronic contributions to $a_\mu$: of vacuum polarization (HVP) and Hadronic light-by-light (HLBL) type. The blob represents all QCD interactions among the quark and anti-quark, dominated by the low-energy hadronic effects.}
\centering
\end{figure}

In Section \ref{sec_RchiT} we give a brief summary of the main features of Resonance Chiral Theory (see Ref.~\cite{ReviewRChT} for a recent review), whose predictions for $a_\mu^{\rm HVP,\,HLBL}$ will be overviewed and compared to other results, in Sections \ref{sec_HVP} and \ref{sec_HLBL}, respectively. We will conclude and give our outlook in Section \ref{sec_Concl}.

\section{A brief introduction to Resonance Chiral Theory}~\label{sec_RchiT}
Quantum Chromodynamics (QCD), the $SU(3)$ gauge theory based on the color degree of freedom~\cite{Fritzsch:1973pi}, predicts -for the six known quarks- that the quantum evolution of the intensity of their interactions is such that the asymptotic freedom of the (anti)quarks and gluons in the ultraviolet corresponds to infrared slavery at very low energies~\cite{Gross:1973id, Politzer:1973fx}. In particular, this prevents perturbative computations in the sub-GeV regime. Instead, one can either study QCD within the lattice~\cite{Wilson:1974sk}, or use effective Lagrangians~\cite{Weinberg:1978kz} that incorporate the symmetries of the fundamental theory using more convenient degrees of freedom, which render perturbative computations again possible.

Within the latter option, Chiral Perturbation Theory ($\chi PT$) emerged~\cite{Gasser:1983yg,Gasser:1984gg}, as dual to QCD at low enough energies, explaining pion, kaon and eta meson physics precisely (including the chiral anomaly~\cite{Wess:1971yu,Witten:1983tw}). However, as one approaches the masses of the $\rho(770)$ meson and other resonances, going to higher orders in $\chi PT$~\cite{Bijnens:1999sh, Bijnens:2001bb} is no longer effective, and other unitarization procedures need to be considered.

Resonance Chiral Theory ($R\chi T$)~\cite{Ecker:1988te,Ecker:1989yg} extends the energy range of applicability of $\chi PT$ beyond the lightest resonance masses, by including them as explicit degrees of freedom in the theory. This is done based on the large number of colors limit of QCD~\cite{tHooft:1973alw,tHooft:1974pnl}, that is equivalent to a semiclassical expansion for meson fields. The theory is built upon chiral symmetry for the lightest pseudo-Goldstone boson particles and unitary symmetry for the resonance fields.

The following two observations are important for understanding that the unknown couplings do not grow out of control. First, the number of resonance fields in the operators is limited by the specific problem under study. Second, there is a balance between the order of the chiral tensors involved and the fulfillment of QCD short-distance constraints that need to be required to ensure formalism-independence of the observables~\cite{Ecker:1988te,Ecker:1989yg}. In this way, the resonance regime smoothly interpolates between the known chiral and short-distance limits of the QCD results. The relations among the $R\chi T$ couplings that are obtained requiring the asymptotic QCD results increase the predictive power of the theory, as will be illustrated in the context of $a_\mu^{\rm HVP,\,HLbL}$ along this review.

We will recall the $R\chi T$ Lagrangian originally derived in Ref.~\cite{Ecker:1988te}. It will serve to illustrate the main ideas underlying the theory, as well as to understand its subsequent extensions~\cite{Cirigliano:2006hb,Kampf:2011ty}, which are also used in the applications we will consider in the next sections.

The $R\chi T$ Lagrangian can be arranged as an expansion in the number of resonance fields, in such a way that the part without them is formally identical to $\chi PT$. However, the values of their low-energy constants (LECs) are different, as resonance exchanges contribute to them~\cite{Ecker:1988te,Ecker:1989yg}. In fact, saturation of the next-to-leading order ($\mathcal{O}(p^4)$ in the chiral counting, in which $p_\pi^2\sim m_\pi^2\sim m_{u,d}$, for two flavors) LECs is achieved, only by the effects of integrating the resonances out~\footnote{Then, the corresponding coefficients of these $\chi PT$ operators, within $R\chi T$, vanish. A corresponding result has not been demonstrated yet at the next order; see Refs.~\cite{Cirigliano:2006hb,Kampf:2011ty}.}, provided that these are described using antisymmetric tensors, as we shall do in the following.

The $R\chi T$ Lagrangian is constructed employing the following chiral tensors, 
\begin{eqnarray}
u_\mu=i\left[u^\dagger\left(\partial_\mu-ir_\mu\right)u-u\left(\partial_\mu-i\ell_\mu\right)u^\dagger\right],\quad \chi_{\pm}=u^\dagger \chi u^\dagger \pm u \chi^\dagger u,\nonumber\\
f_\pm^{\mu\nu}=uF_L^{\mu\nu}u^\dagger\pm u^\dagger F_R^{\mu\nu}u,\quad h_{\mu\nu}=\nabla_\mu u_\nu+\nabla_\nu u_\mu,
\end{eqnarray}
where
\begin{equation} \label{eq:sources}
\chi=2B(s+ip),\quad F_X^{\mu\nu}=\partial^\mu x^\nu - \partial^\nu x^\mu - i [x^\mu,x^\nu]\quad (x=\ell,r),
\end{equation}
and~\cite{Coleman:1969sm,Callan:1969sn}
\begin{equation}\label{eq:defu}
u=\mathrm{exp}\left(i\frac{\Phi}{\sqrt{2}F}\right)\,,
\end{equation}
with $\Phi=\sum_0^8\varphi^a \lambda_a/2$, and
\begin{equation}\label{eq:PhiMatrix}
        \Phi=\begin{pmatrix}
            \frac{1}{\sqrt{2}}(
            \pi^0 +C_q\eta+C_q^{\prime} \eta^{\prime}) & \pi^+& K^+\\
            \pi^-&\frac{1}{\sqrt{2}}(-
            \pi^0 +C_q\eta+C_q^{\prime} \eta^{\prime}) & K^0\\
            K^-& \overline{K}^0& -C_s\eta+C_s^\prime \eta^{\prime}\\
        \end{pmatrix}\,,
\end{equation}
including the large-$N_C$ pion decay constant, $F$, and the $\eta-\eta^\prime$ mixing parameters, $C_{q,s}^{(\prime)}$~\cite{Kaiser:2000gs}. These chiral tensors transform as $X\overset{G}\to h X h^\dagger$, with $h\in H$, where $H,G$ ($G\subset H$) characterize the pattern of spontaneous symmetry breakdown
\begin{equation}\label{eq:SymBreakPattern}
G\equiv SU(3)_L\otimes SU(3)_R \to H\equiv SU(3)_{V=L+R}\,
\end{equation}
in low-energy QCD.

The operators linear in resonance fields (${}^{R}$) and with an $\mathcal{O}(p^2)$ chiral tensor (${}_{(2)}$), which saturate the $\mathcal{O}(p^4)$ $\chi PT$ LECs upon integrating resonances out, are

\begin{equation}\label{eq:originalIntL}\begin{split}
\mathcal{L}^R_{(2)}=\sum_i&\bigg(\frac{F_{V_i}}{2\sqrt{2}}\langle V_i^{\mu\nu}f_{+\mu\nu}\rangle + \frac{i G_{V_i}}{\sqrt{2}}\langle V_i^{\mu\nu} u_\mu u_\nu\rangle + \frac{F_{A_i}}{2\sqrt{2}}\langle A_i^{\mu\nu}f_{-\mu\nu}\rangle\\[0.3ex]
&+c_{d_i} \langle S_i u^\mu u_\mu\rangle + c_{m_i} \langle S_i \chi_+\rangle+i d_{m_i}\langle P_i \chi_- \rangle\,\bigg).
\end{split}\end{equation}

The operators non-linear in resonance fields can we written (using a similar notation)
\begin{equation}\label{eq:extensionsofIntL}
\mathcal{L}^{R,\,\mathrm{even+odd}}_{(4)}+\mathcal{L}^{RR^{(\prime)},\,\mathrm{even+odd}}_{(2)}+\mathcal{L}^{RR^{(\prime)}R^{(\prime \prime)},\,\mathrm{even+odd}}_{(0)}\,,
\end{equation}
where for $R=S,P,V,A$ the operators can be found in Ref.~\cite{Cirigliano:2006hb} (\cite{Kampf:2011ty}) for the even(odd)-intrinsic parity sector. For tensor resonances, the operators can be found in Refs.~\cite{Ecker:2007us, Chen:2023ybr, Estrada:2025bty} (see Sect.~\ref{sec_HLBLTpoles}).

The corresponding couplings are restricted by imposing short-distance QCD constraints. For instance, limiting ourselves to the lowest lying multiplets ($i=1$), the couplings in Eq.~(\ref{eq:originalIntL}) can all be predicted in terms of $F$ and the corresponding resonance masses for the different multiplets (neglecting their flavor symmetry breaking), demanding the QCD high-energy constraints to the two-point correlators in the resonance theory~\cite{Ecker:1988te}. Generalizations of these have been studied, with extended Lagrangians and up to three-point Green functions~\cite{Cirigliano:2006hb, Kampf:2011ty, Roig:2013baa}. Ultimately, the phenomenological success of its applications validate the convergence of the $1/N_C$ expansion in $R\chi T$~\cite{ReviewRChT}.

\section{HVP contributions to \texorpdfstring{$a_\mu$}{Lg}}\label{sec_HVP}
We will summarize in this section the results obtained for $a_\mu^\mathrm{HVP,\,LO}$, either using tau decay (Sect.~\ref{sec_HVPTaus}) or $e^+e^-$ cross-section input (Sect.~\ref{sec_HVPee}), employing $R\chi T$. Noteworthy, in the WP2~\cite{Aliberti:2025beg} (WP25), the tau-based result was the only data-driven prediction quoted for $a_\mu^{\rm HVP}$, due to the incompatibility between the $e^+e^-$ experiments for the fundamental $\pi^+\pi^-$ channel. Both groups of results will be compared to the WP2 lattice QCD prediction, acknowledged as the SM one in Ref.~\cite{Aliberti:2025beg}. A significant difference between the results obtained using $e^+e^-$ or tau input could be explained by new physics affecting only the latter~\cite{Miranda:2018cpf,Cirigliano:2018dyk,Gonzalez-Solis:2020jlh,Cirigliano:2021yto}.

The $a_\mu^\mathrm{HVP,\,LO}$ contribution can be computed, using $e^+e^-\to\mathrm{hadrons}$ data, using~\cite{Brodsky:1967sr}
\begin{equation}\label{eq_amuHVPLO}
a_\mu^\mathrm{HVP,\,LO}=\frac{1}{4\pi^3}\int_{m_{\pi^0}^2}^{\infty}\mathrm{d}sK(s)\sigma^0(e^+e^-\to \mathrm{hadrons})(s)\,,
\end{equation}
with a known kernel, $K(s)\sim 1/s$, that enhances low-energy contributions ($s$ is the hadrons invariant mass squared), in such a way that the $\pi^+\pi^-$ cut gives $\sim73\%$ of the full result (and even a larger part of the uncertainty). The upper-index ${}^0$ specifies that the bare cross-section (removing vacuum polarization effects) is to be integrated.

Given the inconsistency between the $\pi^+\pi^-$ datasets from different experiments (particularly, the old KLOE~\cite{KLOE-2:2017fda}-BaBar~\cite{BaBar:2012bdw} disagreement was exacerbated by the new CMD-3 data~\cite{CMD-3:2023alj,CMD-3:2023rfe}), the WP2 did not provide a number based on this procedure (the different groups doing such combinations~\cite{Colangelo:2018mtw,Davier:2019can,Keshavarzi:2019abf} did not update their results after CMD-3 data appeared, making evident this incompatibility, that is still unresolved; forthcoming data from the flavor factories will be crucial to solve this puzzle).

\subsection{Using tau decay input}\label{sec_HVPTaus}
Tau decays into two pions~\footnote{In principle, this is analogous for other channels probed by vector current, like $\bar{K}K,4\pi,\dots$, although we will restrict ourselves to the dominant case here. Taking the input for these subdominant channels from either source does not affect the results at the current precision.} can be used instead of $\sigma^0(e^+e^-\to \mathrm{hadrons})(s)$ in Eq.~(\ref{eq_amuHVPLO}), thanks to~\cite{Alemany:1997tn,Cirigliano:2001er,Cirigliano:2002pv,Davier:2010fmf}
\begin{equation}\label{eq_eetotau}
\sigma^0(e^+e^-\to \pi^+\pi^-)(s)=\frac{K_\sigma(s)}{K_\Gamma(s)}\frac{\mathrm{d}\Gamma(\tau^-\to\pi^-\pi^0\nu_\tau[\gamma])}{\mathrm{d}s}\frac{R_\mathrm{IB}(s)}{S_\mathrm{EW}}\,.
\end{equation}
In Eq.~(\ref{eq_eetotau}), the first factor involves only masses and fundamental couplings, that are known with extreme precision. The second factor is the tau decay rate, measured more precisely at ALEPH~\cite{ALEPH:2005qgp,Davier:2013sfa} (the best branching fraction) and at Belle~\cite{Belle:2008xpe} (the most accurate normalized spectrum), and also at CLEO~\cite{CLEO:1999dln} and OPAL~\cite{OPAL:1998rrm}. The results of these experiments are compatible within their uncertainties, which are however larger than those of the $e^+e^-$ measurements. The last factor has, as numerator,
\begin{equation}
R_\mathrm{IB}(s)=\frac{\mathrm{FSR}(s)}{G_\mathrm{EM}(s)}\frac{\beta^3_{\pi^+\pi^-}}{\beta^3_{\pi^-\pi^0}}\Bigg|\frac{F_V(s)}{f_+(s)}\Bigg|^2,
\end{equation}
including final-state radiation (FSR), long-distance (encoded in the $G_\mathrm{EM}$ function), phase-space (ratio of $\beta^3$ functions) and neutral-to-charged current form factor corrections (last factor). The last denominator of Eq.~(\ref{eq_eetotau}) is the short-distance and universal electroweak radiative correction~\cite{Marciano:1988vm,Braaten:1990ef,Erler:2002mv,Cirigliano:2023fnz}. We give more details about these in the following, anticipating that the $G_\mathrm{EM}$ function and the form factor correction are the most challenging effects to compute accurately nowadays.

The phase-space correction is in principle trivial, although its precise value will depend slightly on the treatment of the data (binning, for instance) to which it is applied. The scalar QED computation of FSR is accurate enough for present purposes. In the WP2~\cite{Aliberti:2025beg}, it was first published that there is an uncertainty concerning the scheme-dependence cancellation between the $G_\mathrm{EM}$ and $S_\mathrm{EW}$ (first noted by Vincenzo Cirigliano) that has a considerable uncertainty in the IB correction to $a_\mu$. This shall be pushed to the next order with an improved calculation of the $G_\mathrm{EM}$, including all long-distance QED effects up to $\mathcal{O}(\alpha)$.

The $G_\mathrm{EM}$ function~\cite{Cirigliano:2001er, Cirigliano:2002pv} encapsulates long-distance electromagnetic corrections. It includes structure-dependent parts, that were evaluated in these references. using the original $R\chi T$ Lagrangian~\cite{Ecker:1988te}, with the interactions in Eq.~(\ref{eq:originalIntL}). The IB correction thus obtained was $-1.0\times 10^{-10}$, with no uncertainty quoted. In these analyses, using the Guerrero-Pich form factor~\cite{Guerrero:1997ku}, the correction induced by the form factors ratio was $(6.1\pm2.6)\times 10^{-10}$, with $\sim 60\%$ coming from $\rho^0$-$\omega$ mixing and the rest from the charged and neutral width difference (including pion mass dependence and the distinct $\Gamma(\rho\to\pi\pi(\gamma))$ for charged and neutral channels).

Then, a vector meson dominance (VMD) evaluation~\cite{Flores-Baez:2006yiq} found an important role of the $\rho$-$\omega$-$\pi$ vertex (that starts contributing to higher orders in $\chi PT$), which motivated the extension of the original $R\chi T$ analysis, including now the effect of the operators in Refs.~\cite{Cirigliano:2006hb, Kampf:2011ty}, which was performed in Ref.~\cite{Miranda:2020wdg}. This allowed assigning an uncertainty to the original computation in Ref.~\cite{Cirigliano:2002pv}, which was based on the contribution of the operators of the extended Lagrangians which have their coefficients fixed by short-distance QCD constraints (neglecting all others). This, including new data and using the dispersive form factor in Ref.~\cite{GomezDumm:2013sib}, yielded the correction  $(-1.7^{+0.6}_{-1.4})\times10^{-10}$~\footnote{The improved results in Ref.~\cite{Gonzalez-Solis:2019iod} were also employed in later analyses.}, compatible with the VMD result~\cite{Davier:2010fmf}~\footnote{We disregard the result obtained including all operators in Refs.~\cite{Cirigliano:2006hb,Kampf:2011ty}, as its one-sigma uncertainty is $\sim 5.5\times10^{-10}$, falsely augmented by the huge number of free couplings.}. In the previous number, the virtual structure-dependent long-distance contributions were only estimated, based on the results of Refs.~\cite{Arroyo-Urena:2021nil, Arroyo-Urena:2021dfe} for the one-meson modes (see also Ref.~\cite{Escribano:2023seb}).

$R\chi T$ results~\cite{Masjuan:2023qsp} were found to agree with lattice QCD~\cite{RBC:2018dos,RBC:2023pvn,Borsanyi:2020mff,Ce:2022kxy,ExtendedTwistedMass:2022jpw,Giusti:2021dvd} in the most commonly used Euclidean-time windows. CMD-3 measurements~\cite{CMD-3:2023rfe,CMD-3:2023alj} agree with both, and BaBar~\cite{BaBar:2012bdw} was compatible, while a significant tension with KLOE~\cite{KLOE-2:2017fda} was manifest. In light of this, Ref.~\cite{Castro:2024prg} reconsidered the IB form factor correction, using diverse inputs. Among them, analyticity tests favored the dispersive one (based on Ref.~\cite{Colangelo:2018mtw}) and Gounaris-Sakurai~\cite{Gounaris:1968mw}, GS, (used to gauge the model-dependence). This yields the correction $(+1.5\pm1.7)\times10^{-10}$. More recently, Ref.~\cite{Davier:2025jiq} used GS and Kühn-Santamaría \cite{Kuhn:1990ad} fits to extract this correction from data, with a compatible  result (albeit with larger uncertainty). The precision of current data does not seem to allow for a similar procedure using a dispersive formalism.

Finally, the WP2~\cite{Aliberti:2025beg} noticed that present data are still compatible with an IB correction to the $\rho$-$\pi$-$\pi$ coupling that might dominate the overall uncertainty in the tau-based $a_\mu^\mathrm{HVP,\,LO}$ value. This, however, was obtained estimating the corresponding LECs as an $\mathcal{O}(2\alpha/\pi)$ effect, that looks overly conservative. According to us, half of this uncertainty would still be a conservative estimate, rendering the $\tau$-based as the most precise determination of $a_\mu^\mathrm{HVP,\,LO}$. These results are summarized in Table \ref{tab:summary_tau_IB}. There are ongoing efforts to make these predictions more precise, as well as interesting developments in the lattice QCD and dispersive approaches~\cite{Aliberti:2025beg,Bruno:2018ono} that would yield independent predictions for these corrections in the future.

\begin{table}[t]
\footnotesize
	\centering	\renewcommand{\arraystretch}{1.3}
	\begin{tabular}{lrrrrr}
	\toprule
	& &  Refs.~\cite{Davier:2023fpl,Davier:2010fmf} & Ref.~\cite{Castro:2024prg} & WP2 estimate~\cite{Aliberti:2025beg}\\\midrule
    Phase space  & & $-7.88$ & $-7.52$ &  $-7.7(2)$ \\
    $S_{\text{EW}}$ & & $-12.21(15)$  & $-12.16(15)$ &    $-12.2(1.3)$ \\
    $G_\text{EM}$ & & $-1.92(90)$ & $-1.67^{+0.60}_{-1.39}$ &  $-2.0(1.4)$ \\
    FSR   & &  $4.67(47)$ &  $4.62(46)$ & $4.5 (3)$ \\
    $\rho$--$\omega$ mixing & &  $4.0(4)$ & $2.87(8)$ &   $3.9(3)$
   \\\midrule 
     &  $ \Delta M_\rho  $  &     $0.20(^{+27} _{-19})(9)$    & $1.95^{+1.56}_{-1.55}$ &     \\
      &  $\Delta \Gamma_\rho ( \Delta M_\pi )  $  &      $4.09(0)(7)$      & $3.37$ &    \\
            $\frac{F_\pi^V}{f_+}$ (w/o $\rho$--$\omega$)     &  $ \Delta \Gamma_\rho (\pi \pi \gamma)$  &      $ -5.91(59)(48)$      & $-6.66(73)$ &    $-1.5(4.7)$ \\
                              &  $ \Delta \Gamma_\rho ( g_{\rho \pi \pi} )$  &     --     & -- &       \\
    &   Total     &      $-1.62(65)(63)$      & $-1.34^{+1.72}_{-1.71}$ &   
    \\\midrule
    Sum & & $-14.9(1.9)$ &  
       $-15.20^{+2.26}_{-2.63}$ &
  
     $-15.0 (5.1)$
    \\
	\bottomrule
	\renewcommand{\arraystretch}{1.0}
	\end{tabular}
	\caption{Summary of the different classes of IB corrections contributing to 
    $\Delta a_\mu^{\text{HVP, LO}}[\pi\pi,\tau]$
    (in units of $10^{-10}$). For the $\rho$--$\omega$ mixing, a correction of $4\times10^{-10}$ (with $\lesssim10\%$ uncertainty) is supported by the very precise dispersive analyses of Refs.~\cite{Colangelo:2022prz,Hoferichter:2023sli}, a result which is also found in the GS fits of Ref.~\cite{Castro:2024prg} (in this case, the form factor corrections gets shifted, so that the sum presented in this table gets barely affected).}
 \label{tab:summary_tau_IB}
\end{table}

Using the IB correction in Table~\ref{tab:summary_tau_IB}, it is found that 
\begin{equation}\label{eq_HVPTau}
  a_\mu^{\mathrm{HVP,\,LO}}[\pi\pi,\tau]=517.2(2.8)_{\rm exp}(5.1)_{\rm th}
  ,
\end{equation}\\
which is $1.7\sigma$ larger that the WP1 result (WP20), $a_\mu^{\mathrm{HVP,\,LO}}[\pi\pi,e^+e^-]=506.0(2.8)_{\rm exp}
$\cite{Aoyama:2020ynm}. Taking the remaining channels from $e^+e^-$ data \cite{Aoyama:2020ynm}, results in
\begin{equation}\label{eq_tauresult}
  a_\mu^{\mathrm{HVP,\,LO}}[\tau]=704.5(6.2)
  ,
\end{equation}
$1.5\sigma$ larger that the WP1 number~\cite{Aoyama:2020ynm}: $a_\mu^{\mathrm{HVP,\,LO}}[\rm WP20]=693.1(4.0)
$~\cite{Aoyama:2020ynm} and consistent at $1.0\sigma$ with the WP2 result, $a_\mu^{\mathrm{HVP,\,LO}}[\rm WP25]=713.2(6.1)
$~\cite{Aliberti:2025beg}. Using $\tau$ input, the agreement of the current world average and the theory prediction would be slightly reduced, from $0.6$ to $1.3\sigma$.  With the error of the form factor correction halved, the theoretical uncertainty in Eq.~(\ref{eq_HVPTau}) would be $3.1$ units and the experiment-SM discrepancy would be around $2\sigma$.

A Belle-II measurement of the di-pion tau decay, as well of the process with a real photon radiated (for which no data exist at the moment) will be crucial in reducing the uncertainty of this tau-based prediction, and keep competitivity with the ever-increasing precision of the $a_\mu^{\rm HVP}$ lattice calculations.\\\\

\subsection{Using \texorpdfstring{$\sigma(e^+e^-\to\rm hadrons)$}{Lg} input}\label{sec_HVPee}
In this subsection, we review the results obtained in Refs.~\cite{Qin:2020udp,Wang:2023njt, Qin:2024ulb}, for the evaluation of the most important contributions to $a_\mu^\mathrm{HVP,\,LO}$ using $e^+e^-$ data to fit $R\chi T$ descriptions of the relevant vector form factors for the different relevant channels. We quote in the following their most updated result (we omit the factor $\times10^{-10}$) for every computed mode, integrating up to $2.3$ GeV the fitted data, using Eq.~(\ref{eq_amuHVPLO}):
\begin{eqnarray}
  a_\mu^{\pi\pi}=505.89\pm2.34\,,\quad a_\mu^{KK}&=&23.03\pm0.43\,,\quad  a_\mu^{K_L^0K_S^0}=12.67\pm0.24\,,\nonumber\\
  a_\mu^{3\pi}=48.68\pm1.45\,,\;\;\quad a_\mu^{\pi\pi\eta}&=&1.49\pm0.12\,,\;\;\;\quad a_\mu^{K\bar{K}\pi}=3.18\pm0.07\,,\nonumber\\
  a_\mu^{\pi^0\gamma}=4.61\pm0.10 \,,\;\;\;\;\;\,\quad a_\mu^{\eta\gamma}&=&0.66\pm0.06 \,.
\end{eqnarray}
In this way, including other modes from the WP1\cite{Aoyama:2020ynm}, the result
\begin{equation}\label{eq_eeresult}
a_\mu^\mathrm{HVP,\,LO}= (694.1\pm3.1)\times10^{-10}
\end{equation}
is found. This agrees very well with $a_\mu^{\mathrm{HVP,\,LO}}[\rm WP20]$ quoted above. As such, it is in some tension with the WP2 number ($2.8\sigma$) and departs slightly with respect to the tau-data based prediction, Eq.~(\ref{eq_tauresult}) ($1.5\sigma$). Eq.~(\ref{eq_eeresult}) would imply a difference between the world average of $a_\mu$ measurements and Eq.~(\ref{eq_eeresult}) of $6.6\sigma$. A reanalysis of the $\pi\pi$ result above would be in order once the landscape of $e^+e^-$ data clears up.

\section{HLBL contributions to \texorpdfstring{$a_\mu$}{Lg}}\label{sec_HLBL}
We will recap in this section the different results obtained, within $R\chi T$, for the diverse pieces conforming the HLBL part at LO and compare them to other results. These are, the pseudoscalar pole (Sec.~\ref{sec_HLBLPpoles}) and box contributions (Sec.~\ref{sec_HLBLPboxes}), as well as the axial (Sec.~\ref{sec_HLBLApoles}), scalar (Sec.~\ref{sec_HLBLSpoles}) and tensor pole (Sec.~\ref{sec_HLBLTpoles}) pieces, that will be complemented with a concise discussion on short-distance contraints and other contributions (Sec.~\ref{sec_HLBLSD}), as well as on the full $a_\mu^{\rm HLBL}$ obtained using $R\chi T$ input.

\subsection{Pseudoscalar pole contributions}\label{sec_HLBLPpoles}
The pseudoscalar pole contribution to the muon g-2, $a_\mu^\mathrm{P-pole,\,HLBL}$, basically saturates the whole HLbL result and is given as an integral~\cite{Knecht:2001qf}
\begin{equation}
a_\mu^{\rm P-pole}=-\frac{2\alpha^3}{3\pi^2}\int_0^\infty \mathrm{d}Q_1 \mathrm{d}Q_2 \int_0^1 \mathrm{d}t \sqrt{1-t^2}Q_1^3 Q_2^3 \left[F_1 P_6 I_1(Q_1,Q_2,t)+F_2 P_7 I_2(Q_1,Q_2,t)\right]\,,
\end{equation}
which only needs the transition form factor, $\mathcal{F}$, as hadron input~\footnote{$P_{6,7}$ are time-like pion propagators of momenta $Q_{2,3}^2$, $Q_i=|Q_i|,Q_3^2=Q_1^2+Q_2^2+2Q_1^2Q_2^2t$ and the $I_i=I_i(Q_1^2,Q_2^2,t)$ are known functions.}, via
\begin{equation}
    F_1=\mathcal{F}_{P\gamma^*\gamma^*}(Q_1^2,Q_3^2)\mathcal{F}_{P\gamma^*\gamma^*}(Q_2^2,0)\,,\quad F_2=\mathcal{F}_{P\gamma^*\gamma^*}(Q_1^2,Q_2^2)\mathcal{F}_{P\gamma^*\gamma^*}(Q_3^2,0)\,.
\end{equation}
The knowledge of these form factor is very good, thanks to their high-quality measurements \cite{ParticleDataGroup:2024cfk,BaBar:2009rrj,BaBar:2011nrp,Belle:2012wwz,CELLO:1990klc,CLEO:1997fho,L3:1997ocz,BaBar:2018zpn}. In this respect, recent lattice results for double virtuality~\cite{Gerardin:2023naa,ExtendedTwistedMass:2022ofm,ExtendedTwistedMass:2023hin} are extremely valuable, since there are almost no data for this configuration. In addition to this precious information, the following properties are known from QCD:
\begin{itemize}
\item The chiral limit: $\mathcal{F}_{\pi^0\gamma\gamma}(0,0)=-\frac{N_C}{12\pi^2F}$. Up to chiral symmetry breaking corrections, the normalization of these $P$ form factors (to be taken into account in the two items below) are related as $1:(5C_q-\sqrt{2}C_s)/3:(5C_q^\prime+\sqrt{2}C_s^\prime)/3$ ($\pi^0:\eta:\eta'$), with $C_{q,s}^{(\prime)}$ parametrizing the $\eta-\eta^\prime$ mixing~\cite{Kaiser:2000gs,Feldmann:1998vh,Feldmann:1998sh}.
\item The Brodsky-Lepage limit: $\lim_{Q^2\to\infty}-Q^2\mathcal{F}_{\pi^0\gamma^{(*)}\gamma^{(*)}}(-Q^2,0)=2F$~\cite{Brodsky:1973kr,Lepage:1980fj}.
\item The double virtuality limit: $\lim_{Q^2\to\infty} -Q^2 \mathcal{F}_{\pi^0\gamma^{(*)}\gamma^{(*)}}(Q^2,Q^2)=\frac{2F}{3}$~\cite{Nesterenko:1983ef,Novikov:1983jt}.
\end{itemize}
The last two get corrections from the anomalous dimension of the singlet axial current~\cite{Leutwyler:1997yr}, although current data are not yet sensitive to these effects. Subleading perturbative corrections to the last limit were computed using the operator product expansion (OPE) of QCD and its coefficient was determined from QCD sum rules~\cite{Novikov:1983jt}.

There have been several studies of the $\mathcal{F}_{P\gamma^{(*)}\gamma^{(*)}}(Q_i^2,Q_j^2)$ and their corresponding contributions to $a_\mu$. Here we will focus on Ref.~\cite{Estrada:2024cfy} (see also the earlier $R\chi T$ studies \cite{Kampf:2011ty,Czyz:2012nq,Roig:2014uja,Guevara:2018rhj,Kadavy:2022scu}), that was used to obtain $a_\mu^\mathrm{SM}$ in the WP2~\cite{Aliberti:2025beg}.

This analysis was capable of reproducing the three known limits and accounted for their relevant corrections, as well as provided good fits to all data (notably including the lattice results for double virtuality)~\footnote{See Figs. 3-7 in Ref.~\cite{Estrada:2024cfy}.}, in a setting with two vector and two pseudoscalar multiplets of resonances, using the extended Lagrangian in Refs.~\cite{Ecker:1988te, Cirigliano:2006hb, Kampf:2011ty} and their needed generalizations. The most important outcome of Ref.~\cite{Estrada:2024cfy} is ($10^{-10}$ is omitted below)
\begin{align}
a_{\mu}^{ \{\pi^0,\eta,\eta' \}-\rm poles } &=  \Big\{
6.19(0.06)(^{+0.24}_{-0.15})\,,
1.52(0.05)(^{+0.11}_{-0.08})\,,
1.42(0.07)(^{+0.14}_{-0.09})
\Big\}
\,, \notag\\
a_\mu^\text{P-poles}&=9.13(0.10)(^{+0.30}_{-0.19})
\label{eq_PpolesRChT}
\, , 
\end{align}
where the first(second) uncertainty is statistical(systematic). The latter is dominated by cutting the infinite resonance multiplets to two, with the heaviest one corresponding to an effective resummation of the effects of the remaining modes. Next in importance are the $1/N_C$ corrections, followed by those induced by combining experimental and lattice data (versus keeping only measurements). Their errors for the three channels ($\lbrace\pi^0,\eta,\eta^\prime\rbrace$) are, in turn, $\lbrace+0.18,\pm0.15,+0.04\rbrace,$ $\lbrace+0.10,\pm0.05,-0.06\rbrace,$ $\lbrace+0.14,\pm0.03,-0.08\rbrace$. Other uncertainties are negligible.
This result agrees nicely with the WP2 values~\cite{Hoferichter:2018dmo,Hoferichter:2018kwz,Holz:2024lom,Holz:2024diw}:
\begin{align}
a_{\mu}^{ \{\pi^0,\eta,\eta' \}-\rm poles } &=  \Big\{
6.30(^{+0.27}_{-0.21})\,,
1.47(0.09)\,,
1.35(0.07)
\Big\}
\,, \notag\\
a_\mu^\text{P-poles}&=9.12(^{+0.29}_{-0.24})\, . \label{eq_P-polesWP2}
\end{align}
Independently, Canterbury approximants~\cite{Masjuan:2017tvw}, holographic QCD~\cite{Cappiello:2019hwh, Leutgeb:2021mpu} and functional methods~\cite{Eichmann:2019tjk,Raya:2019dnh} agree with these results, confirming a very solid knowledge of this dominant contribution to HLBL, that could still be improved with better data on the transition form factors, particularly for double virtuality~\footnote{By the time of submitting this review, ref.~\cite{Zhang:2025ijd} appeared. They extend the study of ref.~\cite{Estrada:2024cfy} by including an additional multiplet of vector resonances and considering also time-like data in the fits. Their results, ${6.18\pm0.18,1.52\pm0.12,1.60\pm0.11;9.28\pm0.23}$, agree well with eqs.~(\ref{eq_PpolesRChT}) and (\ref{eq_P-polesWP2}) for the $\pi^0,\eta$ contributions but are larger for the $\eta^ \prime$, which turns out to be larger than for the $\eta$, which was unexpected.}.\\

\subsection{Pseudoscalar box contributions}\label{sec_HLBLPboxes}
$\pi$ and $K$ box contributions to $a_\mu$ can be computed by knowing the corresponding electromagnetic form factor, $F_{\pi,K}(Q^2)$, by means of~\cite{Colangelo:2017fiz}
\begin{equation}
a_\mu^{\rm P-box}=\frac{\alpha^3}{432\pi^2}\int_0^\infty \mathrm{d}Q_1 \mathrm{d}Q_2 \int_0^1 \mathrm{d}t \sum_i^{12}T_i(Q_1,Q_2,t)\overline{\Pi}_i^{\rm P-box}\,,
\end{equation}
where 
\begin{equation}
\overline{\Pi}_i^{\rm P-box}(Q_1^2,Q_2^2,Q_3^2)=F_P(Q_1^2)F_P(Q_2^2)F_P(Q_3^2)\frac{1}{16\pi^2}\int_0^1\mathrm{d}x\int_0^{1-x}\mathrm{d}yI_i(x,y)\,,
\end{equation}
with known scalar functions $T_i$ and $I_i$~\cite{Colangelo:2017fiz}.

The required form factors have been evaluated, using a dispersive framework, with an R$\chi$T phaseshift seed, in Refs.~\cite{Pich:2001pj,GomezDumm:2013sib,Gonzalez-Solis:2019iod}. With this input, we obtain
\begin{equation}\label{eq_PboxesRChT}
a_\mu^{\pi-\rm box}=-15.8(3)\times10^{-11}\,,\quad a_\mu^{K-\rm box}=-0.51(3)\times10^{-11}\,,
\end{equation}
which compares well with the WP2 numbers~\cite{Aliberti:2025beg}, taken from
Refs.~\cite{Colangelo:2017fiz,Stamen:2022uqh},
\begin{equation}\label{eq_PboxesWP2}
a_\mu^{\pi-\rm box}=-15.9(2)\times10^{-11}\,,\quad a_\mu^{K-\rm box}=-0.48(2)\times10^{-11}\,,
\end{equation}
These results are also supported by Dyson-Schwinger evaluations~\cite{Eichmann:2019bqf, Miramontes:2021exi}.

We do not consider here other smaller box contributions, like those coming from excited pseudoscalars~\cite{Miramontes:2024fgo} and baryons~\cite{Estrada:2024rfi}. 
\subsection{Axial pole contributions}\label{sec_HLBLApoles}
The contribution of axial-vector resonances to $a_\mu$ gained much interest after the celebrated Melnikov-Vainshtein work~\cite{Melnikov:2003xd}, which put forward that all short-distance QCD constraints relevant for the HLBL could not be satisfied only with pseudoscalar exchanges, which could be healed by the axial contributions. The importance of Ref.~\cite{Roig:2019reh}, which evaluated $a_\mu^{\rm axial-poles}$ within $R\chi T$, lies more in clarifying some issues about bases and satisfaction of high-energy constraints (see also e. g. ref.~\cite{Miranda:2021lhb}) than in the actual number that was obtained, as we will explain.

A very important result of this paper is the derivation of a dictionary to translate between the four existing bases~\cite{Pascalutsa:2012pr,Jegerlehner:2015stw,Knecht:2003xy} (two different helicity bases, one quark-model inspired and another one first used in the study of the $<$$VVA$$>$ Green's function) in the literature that were used to compute $a_\mu^{\rm axials}$ previously~\footnote{Later on, a different basis, optimized for the dispersive analysis, was proposed~\cite{Hoferichter:2024fsj,Hoferichter:2024bae}.}, which is non-trivial, given the off-shellness. This happens because of the Landau-Yang theorem~\cite{Landau:1948kw,Yang:1950rg}, whose fulfillment in the Melnikov-Vainshtein analysis~\cite{Melnikov:2003xd} was shown in Ref.~\cite{Roig:2019reh}, contrary to the claims in Refs.~\cite{Pauk:2014rta, Jegerlehner:2015stw}.

This work~\cite{Roig:2019reh} also emphasized the role of axial-vector mesons in saturating the short-distance anomaly constraints, which became a hot topic for the g-2 community since then. However, working at leading order in $1/N_C$, $R\chi T$ only contributes to antisymmetric form factors, which give suppressed contributions to $a_\mu$. Even including the effects of higher orders in the chiral and large-$N_C$ expansions, this works finds a result of
\begin{equation}
a_\mu^{(a_1+f_1+f_1')\rm-poles}=(0.8^{+4.1}_{-0.8})\times10^{-11}\,,
\end{equation}
which is clearly smaller than the WP2 number~\cite{Aliberti:2025beg}
\begin{equation}
a_\mu^{(a_1+f_1+f_1')\rm-poles}=(12.2\pm 4.3)\times10^{-11}\,,
\end{equation}
based on the dispersive analysis~\cite{Hoferichter:2024bae}, that is supported by those of holographic QCD~\cite{Cappiello:2019hwh,Leutgeb:2019gbz}, Regge theory~\cite{Masjuan:2020jsf} and Dyson-Schwinger equations~\cite{Eichmann:2024glq}. Improved measurements of the axial transition form factors would be instrumental in reducing the above uncertainty. 

Given the preceding explanations, it seems that an $R\chi T$ evaluation of these contributions would need to be constructed on the basis of the continuation to the timelike of the successful description of the axial current coupling to three-pion states in the spacelike~\cite{Dumm:2009va,Nugent:2013hxa}, which would be challenging anyway. 

\subsection{Scalar pole contributions}\label{sec_HLBLSpoles}
In principle, the scalar pole contributions to $a_\mu^\mathrm{HLBL}$ could be evaluated using the master integrals derived in Ref.~\cite{Knecht:2018sci}, which can be computed knowing the two form factors that appear in the general decomposition of the $<$$SVV$$>$ Green function. These have been computed in $R\chi T$ in Ref.~\cite{Dai:2019lmj}, using the Lagrangian including the interactions in Ref.~\cite{Cirigliano:2006hb}. Upon imposing short-distance QCD constraints, one form factor depends on a single unknown coupling. The other one, however, depends on seven different (combinations of) couplings. With current knowledge on this sector of the Lagrangian it is impossible to comply with all known high-energy QCD constraints while keeping predictive power~\footnote{Ref.~\cite{Knecht:2018sci} used VMD form factors (violating QCD short-distance behaviour), to find $a_\mu^{\rm a_0+f_0+f_0'}=(-3\pm2)\times10^{-11}$.}.

In the WP2~\cite{Aliberti:2025beg}, the contribution of the lightest degrees of freedom in the scalar sector are split as
\begin{equation}\label{eq_WP2Swaves}
a_\mu^{\rm S-waves}=-9.1(1.0)\times10^{-11}\,,
\end{equation}
including the $I=0,1,2$ $\pi\pi,\bar{K}K,\pi\eta$ rescattering~\cite{Colangelo:2017fiz,Colangelo:2017qdm,Danilkin:2021icn}. Therein, the dominant $I=0$ component can be understood as the $f_0(500)$ contribution, in terms of two-pion states, within the dispersive framework.

\subsection{Tensor pole contributions}\label{sec_HLBLTpoles}
Tensor meson pole contributions to $a_\mu$ have been an area of hot debate recently. First, the holographic QCD analysis of Ref.~\cite{Colangelo:2024xfh} was basically in agreement with earlier results~\cite{Pauk:2014rta, Danilkin:2016hnh}, with the dominant single-tensor meson contribution being $a_\mu^{\rm f_2-pole}\in [0.5,0.8]\times10^{-11}$. Later on, the dispersive group, using a quark-model form factor, obtained $a_\mu^{\rm (f_2+a_2+f_2')-poles}=-2.5(3)\times10^{-11}$~\cite{Hoferichter:2024bae}, with the sign switched with respect to earlier results and a larger magnitude. Later analysis within holographic QCD~\cite{Cappiello:2025fyf,Mager:2025pvz} agree in the increased absolute value, but again find a positive result, caused by the contribution of the $\mathcal{F}_3$ form factor, which was neglected in Ref.~\cite{Hoferichter:2024bae} (that only had the $\mathcal{F}_1$ contribution). This conundrum resulted in a large uncertainty for the WP2 result,
\begin{equation}\label{eq_TpolesWP2}
a_\mu^{\rm (f_2+a_2+f_2')-poles}=(-1\pm4)\times10^{-11}\,.
\end{equation}

The recent $R\chi T$ analysis~\cite{Estrada:2025bty} reaffirms the results obtained by the holographic QCD analysis of Ref.~\cite{Cappiello:2025fyf}, particularly when only $\mathcal{F}_1$ contributes. This study considered at first the minimal Lagrangian (without derivatives) which could ensure a set of short-distance QCD constraints in the singly and doubly virtual asymptotic limits. Within this setting, only $\mathcal{F}_1$ is non-vanishing, yielding

\begin{subequations}\label{eq_T-poles}
    \begin{equation}
        a_\mu^{\rm a_2-pole}=-1.02(10)_{\rm stat}(^{+0.00}_{-0.12})_{\rm syst}\times 10^{-11}, 
        \end{equation}
    \begin{equation}
        a_\mu^{\rm f_2-pole}=-3.2(3)_{\rm stat}(^{+0.4}_{-0.0})_{\rm syst}\times 10^{-11},
    \end{equation} 
    \begin{equation}
        a_\mu^{\rm f_2^\prime-pole}=-0.042(13)_{\rm stat}(^{+0.008}_{-0.000})_{\rm syst}\times 10^{-11},
    \end{equation}
    \begin{equation}
        a_\mu^{(a_2+f_2+f_2^\prime) \rm - pole}=\left(-4.3^{+0.5}_{-0.3}\right) \times 10^{-11}.
    \end{equation}
\end{subequations}

 This last result agrees remarkably with the holographic number~\cite{Cappiello:2025fyf} (when the $\mathcal{F}_3$ contribution is set to zero), $-4.6(6)\times 10^{-11}$, and is larger in modulus than the dispersive value~\cite{Hoferichter:2024bae}, $-2.5(3)\times 10^{-11}$, all in accordance with the WP2 number, Eq.~(\ref{eq_TpolesWP2})~\cite{Aliberti:2025beg}. Allowing for operators with two derivatives that only contribute to $\mathcal{F}_{1,3}$, one can also satisfy some short-distance constraints on $\mathcal{F}_{3}$. In this more general case, the previous results (now dominated by the uncertainty on $\mathcal{F}_3^T(0,0)$) change to 
 \begin{subequations}\label{eq_T-poles_bothFFs}
    \begin{equation}
        a_\mu^{\rm a_2-pole}=+0.47(1.43)_{\rm norm}(0.03)_{\rm stat}(^{+0.00}_{-0.06})_{\rm syst}\times 10^{-11}, 
        \end{equation}
    \begin{equation}
        a_\mu^{\rm f_2-pole}=+1.18(4.18)_{\rm norm}(0.12)_{\rm stat}(^{+0.00}_{-0.24})_{\rm syst}\times 10^{-11},
    \end{equation} 
    \begin{equation}
        a_\mu^{\rm f_2^\prime-pole}=+0.040(0.078)_{\rm norm}(0.002)_{\rm stat}(^{+0.000}_{-0.000})_{\rm syst}\times 10^{-11},
    \end{equation}
    \begin{equation}\label{eq_TpolesRChT}
        a_\mu^{(a_2+f_2+f_2^\prime)\rm - pole}=\left(+1.7\pm 4.4\right) \times 10^{-11},
    \end{equation}
\end{subequations}
that agrees in sign with the holographic QCD outcome, $a_\mu^{a_2+f_2+f_2^\prime \rm - pole}=+3.2(4)\times 10^{-11}$, albeit with a much larger uncertainty. This would be in some tension with the WP2 phenomenological value, Eq.~(\ref{eq_TpolesWP2}), but yield to agreement within a sigma with the lattice result for the entire HLbL contribution, $122.5(9.0)\times 10^{-11}$, Eq.~(\ref{eq_LatticeHLbL}).

Currently, the only measurement by Belle \cite{Belle:2015oin} just probes $\mathcal{F}_1$, which does not help clarify the situation. This will hopefully change with future improved/new data. Double virtuality measurements would constrain $\mathcal{F}_3$, which would be extremely valuable to reduce the uncertainty of this contribution, Eq.~(\ref{eq_TpolesRChT}).

We finally stress that $R\chi T$ generally finds contributions to all five form factors allowed by symmetry (this has been noted for the first time in Ref.~\cite{Estrada:2025bty}, to our knowledge). Computing their contributions to $a_\mu^{\rm HLbL}$ requires a new basis, which is a necessary development of the existing formalism and will motivate future work in this direction.

\subsection{Other contributions}\label{sec_HLBLSD}
Approximately $2/3$ of the whole $a_\mu^\mathrm{HLBL}$ is given by the low-energy contributions, coming from the lightest pseudoscalar poles (Sec.~\ref{sec_HLBLPpoles}) and boxes (Sec.~\ref{sec_HLBLPboxes}), as well as their rescattering contributions (Sec.~\ref{sec_HLBLSpoles}), which are dominated by sub-GeV physics. The remaining parts (including the lightest axial and tensor poles in Sects.~\ref{sec_HLBLApoles} and \ref{sec_HLBLTpoles}) can be classified according to the different energy regions probed.

It has become standard to divide the momentum space between the short-distance regime (with all photon momenta in the HLBL with magnitude larger than $1.5$ GeV), the low-energy contributions (all with modulus smaller than this quantity) and the mixed regime.

The pure short-distance (SD) region is very well-known, by applying repeatedly the OPE to the electromagnetic currents in the HLBL tensor~\cite{Melnikov:2003xd,Bijnens:2019ghy,Bijnens:2020xnl,Bijnens:2021jqo,Bijnens:2022itw,Bijnens:2024jgh}, yielding the result
\begin{equation}\label{eq_SDWP2}
a_\mu^{\rm SD}=\left(6.2^{+0.2}_{-0.3}\right)\times10^{-11}\,.
\end{equation}
The contribution of the charm quark loop uses to be quoted separately,
\begin{equation}\label{eq_cllopWP2}
a_\mu^{\rm c-loop}=3(1)\times10^{-11}\,.
\end{equation}
In the $N_C\to\infty$ limit of QCD an infinite tower of mesons is predicted per set of quantum numbers. Although contributions from the lightest states dominate $a_\mu^\mathrm{HLBL}$, the effects of their excitations are sizable, and accounted for, e. g. within holographic QCD~\cite{Leutgeb:2019gbz,Cappiello:2019hwh,Leutgeb:2021mpu,Leutgeb:2022lqw,Cappiello:2025fyf} or Regge models~\cite{Masjuan:2020jsf}.
The results used in the WP2~\cite{Aliberti:2025beg} for these regions are
\begin{equation}\label{eq_remainingWP2}
a_\mu^{\rm A,S',T,\dots\rm-low}=12.5(5.9)\times10^{-11}\,,\quad a_\mu^{\rm mixed}=15.9(1.7)\times10^{-11}\,.
\end{equation}
We note that the former result agrees extremely well with the WP2 number for the lightest axials and tensors, which also provides $\sim 86\%$ of the quoted uncertainty.

The sum of the different contributions in Eqs.~(\ref{eq_P-polesWP2}), (\ref{eq_PboxesWP2}), (\ref{eq_WP2Swaves}), (\ref{eq_SDWP2}), (\ref{eq_cllopWP2}) and (\ref{eq_remainingWP2}), accounting for an additional uncertainty (of $5.0$ units) in the matching of the different regions, yields the WP2 number
\begin{equation}
a_\mu^\mathrm{HLBL,\,WP2}=103.3(8.8)\times10^{-11}\,.
\end{equation}
Using the $R\chi T$ results for the $P$-poles and -boxes and (more importantly) for the tensors instead, according to Eqs.~(\ref{eq_PpolesRChT}), (\ref{eq_PboxesRChT}) and (\ref{eq_TpolesRChT}) would yield
\begin{equation}
a_\mu^{\mathrm{HLBL,\,with\,}R\chi T\,\rm results}=106.1(9.0)\times10^{-11}\,,
\end{equation}
which is closer to the Lattice QCD number (based on Refs.~\cite{Blum:2019ugy,Chao:2021tvp,Chao:2022xzg,Blum:2023vlm}) in the WP2~\cite{Aliberti:2025beg}
\begin{equation}\label{eq_LatticeHLbL}
a_\mu^\mathrm{HLBL,\, lattice}=122.5(9.0)\times10^{-11}\,.
\end{equation}
\section{Conclusions and outlook}\label{sec_Concl}
We have reviewed the foundations of $R\chi T$ and its use in computing hadronic contributions to $a_\mu$, which benefit from measurements of the relevant form factors, whose prospects for improvement have been underlined.

On the HVP part, the tau-based result, (\ref{eq_tauresult}), is very competitive with the lattice QCD value, both agreeing with the $a_\mu$ measurement and thus placing strong constraints on new physics modifying the SM prediction for the muon g-2. If, instead, $e^+e^-$ data are used, the incompatibility between different data sets (and particularly between KLOE and CMD-3) prevents updating Eq.~(\ref{eq_eeresult}) to include the newest measurements.

With respect to the HLBL contributions, $R\chi T$ obtains very competitive results for the lightest pseudoscalar pole and box. Notably, for the tensor pole contributions, it favors the holographic results, which would require shifting the WP2 number for this contribution. This would imply closer agreement with the lattice $a_\mu^{\rm HLBL}$ result.

\backmatter

\bmhead{Acknowledgements}
E. J. E. is grateful to Conahcyt/Secithi for funding during his Ph. D. A. M. is supported by the European Union’s Horizon 2020 Research and Innovation Programme under grant 824093 (H2020-INFRAIA-2018-1), the Ministerio de Ciencia e Innovación under grant PID2020-112965GB-I00, by the Secretaria d’Universitats i Recerca del Departament d’Empresa i Coneixement de la Generalitat de Catalunya under grant 2021 SGR 00649, by MICINN with funding from European Union NextGenerationEU (PRTR-C17.I1) and by Generalitat de Catalunya. IFAE is partially funded by the CERCA program of the Generalitat de Catalunya. P. R. is partly funded by Conahcyt (México) by project CBF2023-2024-3226, and by MCIN/AEI/10.13039/501100011033 (Spain), grants PID2020-114473GB-I00 and PID2023-146220NB-I00, and by Generalitat Valenciana (Spain), grant PROMETEO2021/071.




\bibliography{sn-bibliography}

@article{RBC:2018dos,
    author = {Blum, T. and Boyle, P. A. and G\"ulpers, V. and Izubuchi, T. and Jin, L. and Jung, C. and J\"uttner, A. and Lehner, C. and Portelli, A. and Tsang, J. T.},
    collaboration = "RBC, UKQCD",
    title = "{Calculation of the hadronic vacuum polarization contribution to the muon anomalous magnetic moment}",
    eprint = "1801.07224",
    archivePrefix = "arXiv",
    primaryClass = "hep-lat",
    doi = "10.1103/PhysRevLett.121.022003",
    journal = "Phys. Rev. Lett.",
    volume = "121",
    number = "2",
    pages = "022003",
    year = "2018"
}

@article{Davier:2025jiq,
    author = "Davier, Michel and Malaescu, Bogdan and Zhang, Zhiqing",
    title = "{Data-based form factor corrections between the two-pion $\tau$ and $e^+e^-$ spectral functions}",
    eprint = "2504.13789",
    archivePrefix = "arXiv",
    primaryClass = "hep-ph",
    month = "4",
    year = "2025"
}

@article{Estrada:2025bty,
    author = "Estrada, Emilio J. and Roig, Pablo",
    title = "{Tensor Meson Pole contributions to the HLbL piece of $a_{\mu}^{\rm{HLbL}}$ within R$\chi$T}",
    eprint = "2504.00448",
    archivePrefix = "arXiv",
    primaryClass = "hep-ph",
    month = "4",
    year = "2025"
}

@article{Marciano:1988vm,
    author = "Marciano, W. J. and Sirlin, A.",
    title = "{Electroweak Radiative Corrections to tau Decay}",
    doi = "10.1103/PhysRevLett.61.1815",
    journal = "Phys. Rev. Lett.",
    volume = "61",
    pages = "1815--1818",
    year = "1988"
}

@article{Braaten:1990ef,
    author = "Braaten, Eric and Li, Chong-Sheng",
    title = "{Electroweak radiative corrections to the semihadronic decay rate of the tau lepton}",
    reportNumber = "NUHEP-TH-90-30",
    doi = "10.1103/PhysRevD.42.3888",
    journal = "Phys. Rev. D",
    volume = "42",
    pages = "3888--3891",
    year = "1990"
}

@article{Erler:2002mv,
    author = "Erler, Jens",
    title = "{Electroweak radiative corrections to semileptonic tau decays}",
    eprint = "hep-ph/0211345",
    archivePrefix = "arXiv",
    journal = "Rev. Mex. Fis.",
    volume = "50",
    pages = "200--202",
    year = "2004"
}

@article{Cirigliano:2023fnz,
    author = "Cirigliano, Vincenzo and Dekens, Wouter and Mereghetti, Emanuele and Tomalak, Oleksandr",
    title = "{Effective field theory for radiative corrections to charged-current processes: Vector coupling}",
    eprint = "2306.03138",
    archivePrefix = "arXiv",
    primaryClass = "hep-ph",
    reportNumber = "LA-UR-22-21034, INT-PUB-23-015",
    doi = "10.1103/PhysRevD.108.053003",
    journal = "Phys. Rev. D",
    volume = "108",
    number = "5",
    pages = "053003",
    year = "2023"
}

@article{Gounaris:1968mw,
    author = "Gounaris, G. J. and Sakurai, J. J.",
    title = "{Finite width corrections to the vector meson dominance prediction for $\rho \to e^+ e^-$}",
    doi = "10.1103/PhysRevLett.21.244",
    journal = "Phys. Rev. Lett.",
    volume = "21",
    pages = "244--247",
    year = "1968"
}

@article{Qin:2020udp,
    author = "Qin, Wen and Dai, Ling-Yun and Portol\'es, Jorge",
    title = "{Two and three pseudoscalar production in e$^{+}$e$^{-}$ annihilation and their contributions to (g-2)$_{\mu}$}",
    eprint = "2011.09618",
    archivePrefix = "arXiv",
    primaryClass = "hep-ph",
    doi = "10.1007/JHEP03(2021)092",
    journal = "JHEP",
    volume = "03",
    pages = "092",
    year = "2021"
}

@article{Kaiser:2000gs,
    author = "Kaiser, Roland and Leutwyler, H.",
    title = "{Large N(c) in chiral perturbation theory}",
    eprint = "hep-ph/0007101",
    archivePrefix = "arXiv",
    reportNumber = "BUTP-00-19",
    doi = "10.1007/s100520000499",
    journal = "Eur. Phys. J. C",
    volume = "17",
    pages = "623--649",
    year = "2000"
}

@article{Feldmann:1998sh,
    author = "Feldmann, T. and Kroll, P. and Stech, B.",
    title = "{Mixing and decay constants of pseudoscalar mesons: The Sequel}",
    eprint = "hep-ph/9812269",
    archivePrefix = "arXiv",
    reportNumber = "WUB-98-40, ADP-98-75-T342, HD-THEP-98-59",
    doi = "10.1016/S0370-2693(99)00085-4",
    journal = "Phys. Lett. B",
    volume = "449",
    pages = "339--346",
    year = "1999"
}

@article{Feldmann:1998vh,
    author = "Feldmann, T. and Kroll, P. and Stech, B.",
    title = "{Mixing and decay constants of pseudoscalar mesons}",
    eprint = "hep-ph/9802409",
    archivePrefix = "arXiv",
    reportNumber = "WU-B-98-2, HD-THEP-98-5",
    doi = "10.1103/PhysRevD.58.114006",
    journal = "Phys. Rev. D",
    volume = "58",
    pages = "114006",
    year = "1998"
}

@article{Novikov:1983jt,
    author = "Novikov, V. A. and Shifman, Mikhail A. and Vainshtein, A. I. and Voloshin, M. B. and Zakharov, Valentin I.",
    title = "{Use and Misuse of QCD Sum Rules, Factorization and Related Topics}",
    reportNumber = "ITEP-71-1983",
    doi = "10.1016/0550-3213(84)90006-3",
    journal = "Nucl. Phys. B",
    volume = "237",
    pages = "525--552",
    year = "1984"
}

@article{Nesterenko:1983ef,
    author = "Nesterenko, V. A. and Radyushkin, A. V.",
    title = "{Local Quark - Hadron Duality and Nucleon Form-factors in {QCD}}",
    reportNumber = "JINR-P2-83-251",
    doi = "10.1016/0370-2693(83)90935-8",
    journal = "Phys. Lett. B",
    volume = "128",
    pages = "439--444",
    year = "1983"
}

@article{BaBar:2009rrj,
    author = "Aubert, Bernard and others",
    collaboration = "BaBar",
    title = "{Measurement of the gamma gamma* ---\ensuremath{>} pi0 transition form factor}",
    eprint = "0905.4778",
    archivePrefix = "arXiv",
    primaryClass = "hep-ex",
    reportNumber = "SLAC-PUB-13641, BABAR-PUB-09-006",
    doi = "10.1103/PhysRevD.80.052002",
    journal = "Phys. Rev. D",
    volume = "80",
    pages = "052002",
    year = "2009"
}

@article{BaBar:2011nrp,
    author = "del Amo Sánchez, P. and others",
    collaboration = "BaBar",
    title = "{Measurement of the $\gamma \gamma^* --> \eta$ and $\gamma \gamma* --> \eta'$ transition form factors}",
    eprint = "1101.1142",
    archivePrefix = "arXiv",
    primaryClass = "hep-ex",
    reportNumber = "SLAC-PUB-14317, BABAR-PUB-10-028",
    doi = "10.1103/PhysRevD.84.052001",
    journal = "Phys. Rev. D",
    volume = "84",
    pages = "052001",
    year = "2011"
}

@article{Belle:2012wwz,
    author = "Uehara, S. and others",
    collaboration = "Belle",
    title = "{Measurement of $\gamma \gamma^* \to \pi^0$ transition form factor at Belle}",
    eprint = "1205.3249",
    archivePrefix = "arXiv",
    primaryClass = "hep-ex",
    reportNumber = "BELLE-PREPRINT-2012-16, KEK-PREPRINT-2012-8",
    doi = "10.1103/PhysRevD.86.092007",
    journal = "Phys. Rev. D",
    volume = "86",
    pages = "092007",
    year = "2012"
}

@article{CELLO:1990klc,
    author = "Behrend, H. J. and others",
    collaboration = "CELLO",
    title = "{A Measurement of the pi0, eta and eta-prime electromagnetic form-factors}",
    reportNumber = "DESY-90-110",
    doi = "10.1007/BF01549692",
    journal = "Z. Phys. C",
    volume = "49",
    pages = "401--410",
    year = "1991"
}

@article{CLEO:1997fho,
    author = "Gronberg, J. and others",
    collaboration = "CLEO",
    title = "{Measurements of the meson - photon transition form-factors of light pseudoscalar mesons at large momentum transfer}",
    eprint = "hep-ex/9707031",
    archivePrefix = "arXiv",
    reportNumber = "SLAC-PUB-9838, CLNS-97-1477, CLEO-97-7",
    doi = "10.1103/PhysRevD.57.33",
    journal = "Phys. Rev. D",
    volume = "57",
    pages = "33--54",
    year = "1998"
}

@article{L3:1997ocz,
    author = "Acciarri, M. and others",
    collaboration = "L3",
    title = "{Measurement of eta-prime (958) formation in two photon collisions at LEP-1}",
    reportNumber = "CERN-PPE-97-110",
    doi = "10.1016/S0370-2693(97)01219-7",
    journal = "Phys. Lett. B",
    volume = "418",
    pages = "399--410",
    year = "1998"
}

@article{BaBar:2018zpn,
    author = "Lees, J. P. and others",
    collaboration = "BaBar",
    title = "{Measurement of the $\gamma^{\star}\gamma^{\star} \to \eta'$ transition form factor}",
    eprint = "1808.08038",
    archivePrefix = "arXiv",
    primaryClass = "hep-ex",
    reportNumber = "BABAR-PUB-18/006, SLAC-PUB-17318",
    doi = "10.1103/PhysRevD.98.112002",
    journal = "Phys. Rev. D",
    volume = "98",
    number = "11",
    pages = "112002",
    year = "2018"
}

@article{Gerardin:2023naa,
    author = "G\'erardin, Antoine and Verplanke, Willem E. A. and Wang, Gen and Fodor, Zoltan and Guenther, Jana N. and Lellouch, Laurent and Szabo, Kalman K. and Varnhorst, Lukas",
    title = "{Lattice calculation of the \ensuremath{\pi}0, \ensuremath{\eta} and \ensuremath{\eta}' transition form factors and the hadronic light-by-light contribution to the muon g-2}",
    eprint = "2305.04570",
    archivePrefix = "arXiv",
    primaryClass = "hep-lat",
    doi = "10.1103/PhysRevD.111.054511",
    journal = "Phys. Rev. D",
    volume = "111",
    number = "5",
    pages = "054511",
    year = "2025"
}

@article{ExtendedTwistedMass:2022ofm,
    author = "Alexandrou, Constantia and others",
    collaboration = "Extended Twisted Mass",
    title = "{\ensuremath{\eta}\textrightarrow{}\ensuremath{\gamma}*\ensuremath{\gamma}* transition form factor and the hadronic light-by-light \ensuremath{\eta}-pole contribution to the muon g-2 from lattice QCD}",
    eprint = "2212.06704",
    archivePrefix = "arXiv",
    primaryClass = "hep-lat",
    doi = "10.1103/PhysRevD.108.054509",
    journal = "Phys. Rev. D",
    volume = "108",
    number = "5",
    pages = "054509",
    year = "2023"
}

@article{ExtendedTwistedMass:2023hin,
    author = "Alexandrou, C. and others",
    collaboration = "Extended Twisted Mass",
    title = "{Pion transition form factor from twisted-mass lattice QCD and the hadronic light-by-light \ensuremath{\pi}0-pole contribution to the muon g-2}",
    eprint = "2308.12458",
    archivePrefix = "arXiv",
    primaryClass = "hep-lat",
    doi = "10.1103/PhysRevD.108.094514",
    journal = "Phys. Rev. D",
    volume = "108",
    number = "9",
    pages = "094514",
    year = "2023"
}

@article{Colangelo:2024xfh,
    author = "Colangelo, Pietro and Giannuzzi, Floriana and Nicotri, Stefano",
    title = "{Hadronic light-by-light scattering contributions to (g-2)\ensuremath{\mu} from axial-vector and tensor mesons in the holographic soft-wall model}",
    eprint = "2402.07579",
    archivePrefix = "arXiv",
    primaryClass = "hep-ph",
    reportNumber = "BARI-TH/24-755",
    doi = "10.1103/PhysRevD.109.094036",
    journal = "Phys. Rev. D",
    volume = "109",
    number = "9",
    pages = "094036",
    year = "2024"
}

@article{Hoferichter:2024bae,
    author = "Hoferichter, Martin and Stoffer, Peter and Zillinger, Maximilian",
    title = "{Dispersion relation for hadronic light-by-light scattering: subleading contributions}",
    eprint = "2412.00178",
    archivePrefix = "arXiv",
    primaryClass = "hep-ph",
    reportNumber = "PSI-PR-24-26, ZU-TH 60/24",
    doi = "10.1007/JHEP02(2025)121",
    journal = "JHEP",
    volume = "02",
    pages = "121",
    year = "2025"
}

@article{Cappiello:2025fyf,
    author = "Cappiello, Luigi and Leutgeb, Josef and Mager, Jonas and Rebhan, Anton",
    title = "{Tensor meson transition form factors in holographic QCD and the muon $g-2$}",
    eprint = "2501.09699",
    archivePrefix = "arXiv",
    primaryClass = "hep-ph",
    month = "1",
    year = "2025"
}

@article{Mager:2025pvz,
    author = "Mager, Jonas and Cappiello, Luigi and Leutgeb, Josef and Rebhan, Anton",
    title = "{Longitudinal short-distance constraints on hadronic light-by-light scattering and tensor meson contributions to the muon $g-2$}",
    eprint = "2501.19293",
    archivePrefix = "arXiv",
    primaryClass = "hep-ph",
    month = "1",
    year = "2025"
}

@article{Hoferichter:2018dmo,
    author = "Hoferichter, Martin and Hoid, Bai-Long and Kubis, Bastian and Leupold, Stefan and Schneider, Sebastian P.",
    title = "{Pion-pole contribution to hadronic light-by-light scattering in the anomalous magnetic moment of the muon}",
    eprint = "1805.01471",
    archivePrefix = "arXiv",
    primaryClass = "hep-ph",
    reportNumber = "INT-PUB-18-016",
    doi = "10.1103/PhysRevLett.121.112002",
    journal = "Phys. Rev. Lett.",
    volume = "121",
    number = "11",
    pages = "112002",
    year = "2018"
}

@article{Hoferichter:2018kwz,
    author = "Hoferichter, Martin and Hoid, Bai-Long and Kubis, Bastian and Leupold, Stefan and Schneider, Sebastian P.",
    title = "{Dispersion relation for hadronic light-by-light scattering: pion pole}",
    eprint = "1808.04823",
    archivePrefix = "arXiv",
    primaryClass = "hep-ph",
    reportNumber = "INT-PUB-18-042",
    doi = "10.1007/JHEP10(2018)141",
    journal = "JHEP",
    volume = "10",
    pages = "141",
    year = "2018"
}

@article{Holz:2024lom,
    author = "Holz, Simon and Hoferichter, Martin and Hoid, Bai-Long and Kubis, Bastian",
    title = "{Precision Evaluation of the \ensuremath{\eta}- and \ensuremath{\eta}'-Pole Contributions to Hadronic Light-by-Light Scattering in the Anomalous Magnetic Moment of the Muon}",
    eprint = "2411.08098",
    archivePrefix = "arXiv",
    primaryClass = "hep-ph",
    doi = "10.1103/PhysRevLett.134.171902",
    journal = "Phys. Rev. Lett.",
    volume = "134",
    number = "17",
    pages = "171902",
    year = "2025"
}

@article{Holz:2024diw,
    author = "Holz, Simon and Hoferichter, Martin and Hoid, Bai-Long and Kubis, Bastian",
    title = "{Dispersion relation for hadronic light-by-light scattering: \ensuremath{\eta} and \ensuremath{\eta}$^{\prime}$ poles}",
    eprint = "2412.16281",
    archivePrefix = "arXiv",
    primaryClass = "hep-ph",
    doi = "10.1007/JHEP04(2025)147",
    journal = "JHEP",
    volume = "04",
    pages = "147",
    year = "2025"
}

@article{Pauk:2014rta,
    author = "Pauk, Vladyslav and Vanderhaeghen, Marc",
    title = "{Single meson contributions to the muon`s anomalous magnetic moment}",
    eprint = "1401.0832",
    archivePrefix = "arXiv",
    primaryClass = "hep-ph",
    doi = "10.1140/epjc/s10052-014-3008-y",
    journal = "Eur. Phys. J. C",
    volume = "74",
    number = "8",
    pages = "3008",
    year = "2014"
}

@article{Masjuan:2020jsf,
    author = "Masjuan, Pere and Roig, Pablo and Sánchez-Puertas, Pablo",
    title = "{The interplay of transverse degrees of freedom and axial-vector mesons with short-distance constraints in g \ensuremath{-} 2}",
    eprint = "2005.11761",
    archivePrefix = "arXiv",
    primaryClass = "hep-ph",
    doi = "10.1088/1361-6471/ac3892",
    journal = "J. Phys. G",
    volume = "49",
    number = "1",
    pages = "015002",
    year = "2022"
}

@article{Eichmann:2024glq,
    author = "Eichmann, Gernot and Fischer, Christian S. and Haeuser, Tim and Regenfelder, Oliver",
    title = "{Axial-vector and scalar contributions to hadronic light-by-light scattering}",
    eprint = "2411.05652",
    archivePrefix = "arXiv",
    primaryClass = "hep-ph",
    doi = "10.1140/epjc/s10052-025-14055-7",
    journal = "Eur. Phys. J. C",
    volume = "85",
    number = "4",
    pages = "445",
    year = "2025"
}

@article{Leutgeb:2019gbz,
    author = "Leutgeb, Josef and Rebhan, Anton",
    title = "{Axial vector transition form factors in holographic QCD and their contribution to the anomalous magnetic moment of the muon}",
    eprint = "1912.01596",
    archivePrefix = "arXiv",
    primaryClass = "hep-ph",
    doi = "10.1103/PhysRevD.101.114015",
    journal = "Phys. Rev. D",
    volume = "101",
    number = "11",
    pages = "114015",
    year = "2020"
}

@article{Raya:2019dnh,
    author = "Raya, Kh\'epani and Bashir, Adnan and Roig, Pablo",
    title = "{Contribution of neutral pseudoscalar mesons to $a_\mu^{HLbL}$ within a Schwinger-Dyson equations approach to QCD}",
    eprint = "1910.05960",
    archivePrefix = "arXiv",
    primaryClass = "hep-ph",
    doi = "10.1103/PhysRevD.101.074021",
    journal = "Phys. Rev. D",
    volume = "101",
    number = "7",
    pages = "074021",
    year = "2020"
}

@article{Leutgeb:2021mpu,
    author = "Leutgeb, Josef and Rebhan, Anton",
    title = "{Hadronic light-by-light contribution to the muon g-2 from holographic QCD with massive pions}",
    eprint = "2108.12345",
    archivePrefix = "arXiv",
    primaryClass = "hep-ph",
    doi = "10.1103/PhysRevD.104.094017",
    journal = "Phys. Rev. D",
    volume = "104",
    number = "9",
    pages = "094017",
    year = "2021"
}

@article{Melnikov:2003xd,
    author = "Melnikov, Kirill and Vainshtein, Arkady",
    title = "{Hadronic light-by-light scattering contribution to the muon anomalous magnetic moment revisited}",
    eprint = "hep-ph/0312226",
    archivePrefix = "arXiv",
    reportNumber = "UH-511-1041-03, FTPI-MINN-03-36, UMN-TH-2224-03",
    doi = "10.1103/PhysRevD.70.113006",
    journal = "Phys. Rev. D",
    volume = "70",
    pages = "113006",
    year = "2004"
}

@article{Belle:2015oin,
    author = "Masuda, M. and others",
    collaboration = "Belle",
    title = "{Study of $\pi^0$ pair production in single-tag two-photon collisions}",
    eprint = "1508.06757",
    archivePrefix = "arXiv",
    primaryClass = "hep-ex",
    reportNumber = "BELLE-PREPRINT-2015-15, KEK-PREPRINT-2015-24",
    doi = "10.1103/PhysRevD.93.032003",
    journal = "Phys. Rev. D",
    volume = "93",
    number = "3",
    pages = "032003",
    year = "2016"
}

@article{Danilkin:2021icn,
    author = "Danilkin, Igor and Hoferichter, Martin and Stoffer, Peter",
    title = "{A dispersive estimate of scalar contributions to hadronic light-by-light scattering}",
    eprint = "2105.01666",
    archivePrefix = "arXiv",
    primaryClass = "hep-ph",
    reportNumber = "INT-PUB-21-11, UWThPh 2021-3",
    doi = "10.1016/j.physletb.2021.136502",
    journal = "Phys. Lett. B",
    volume = "820",
    pages = "136502",
    year = "2021"
}

@article{Colangelo:2017qdm,
    author = "Colangelo, Gilberto and Hoferichter, Martin and Procura, Massimiliano and Stoffer, Peter",
    title = "{Rescattering effects in the hadronic-light-by-light contribution to the anomalous magnetic moment of the muon}",
    eprint = "1701.06554",
    archivePrefix = "arXiv",
    primaryClass = "hep-ph",
    reportNumber = "INT-PUB-17-005, CERN-TH-2017-014, NSF-KITP-17-012",
    doi = "10.1103/PhysRevLett.118.232001",
    journal = "Phys. Rev. Lett.",
    volume = "118",
    number = "23",
    pages = "232001",
    year = "2017"
}

@article{Brodsky:1967sr,
    author = "Brodsky, Stanley J. and De Rafael, Eduardo",
    title = "{SUGGESTED BOSON - LEPTON PAIR COUPLINGS AND THE ANOMALOUS MAGNETIC MOMENT OF THE MUON}",
    reportNumber = "SLAC-PUB-0366",
    doi = "10.1103/PhysRev.168.1620",
    journal = "Phys. Rev.",
    volume = "168",
    pages = "1620--1622",
    year = "1968"
}

@article{Muong-2:2023cdq,
    author = "Aguillard, D. P. and others",
    collaboration = "Muon g-2",
    title = "{Measurement of the Positive Muon Anomalous Magnetic Moment to 0.20~ppm}",
    eprint = "2308.06230",
    archivePrefix = "arXiv",
    primaryClass = "hep-ex",
    reportNumber = "FERMILAB-PUB-23-385-AD-CSAID-PPD",
    doi = "10.1103/PhysRevLett.131.161802",
    journal = "Phys. Rev. Lett.",
    volume = "131",
    number = "16",
    pages = "161802",
    year = "2023"
}

@article{Muong-2:2021ojo,
    author = "Abi, B. and others",
    collaboration = "Muon g-2",
    title = "{Measurement of the Positive Muon Anomalous Magnetic Moment to 0.46 ppm}",
    eprint = "2104.03281",
    archivePrefix = "arXiv",
    primaryClass = "hep-ex",
    reportNumber = "FERMILAB-PUB-21-132-E",
    doi = "10.1103/PhysRevLett.126.141801",
    journal = "Phys. Rev. Lett.",
    volume = "126",
    number = "14",
    pages = "141801",
    year = "2021"
}

@article{Muong-2:2006rrc,
    author = "Bennett, G. W. and others",
    collaboration = "Muon g-2",
    title = "{Final Report of the Muon E821 Anomalous Magnetic Moment Measurement at BNL}",
    eprint = "hep-ex/0602035",
    archivePrefix = "arXiv",
    doi = "10.1103/PhysRevD.73.072003",
    journal = "Phys. Rev. D",
    volume = "73",
    pages = "072003",
    year = "2006"
}

@article{Aliberti:2025beg,
    author = "Aliberti, R. and others",
    title = "{The anomalous magnetic moment of the muon in the Standard Model: an update}",
    eprint = "2505.21476",
    archivePrefix = "arXiv",
    primaryClass = "hep-ph",
    reportNumber = "CERN-TH-2025-101, FERMILAB-PUB-25-0344-T, INT-PUB-25-015,
  IPARCOS-UCM-25-029, KEK Preprint 2025-22, LTH 1403, MITP-25-037, UWThPh
  2025-15, ZU-TH 37/25",
    month = "5",
    year = "2025"
}

@article{Davier:2019can,
    author = "Davier, M. and Hoecker, A. and Malaescu, B. and Zhang, Z.",
    title = "{A new evaluation of the hadronic vacuum polarisation contributions to the muon anomalous magnetic moment and to $\mathbf{\boldsymbol\alpha(m_Z^2)}$}",
    eprint = "1908.00921",
    archivePrefix = "arXiv",
    primaryClass = "hep-ph",
    doi = "10.1140/epjc/s10052-020-7792-2",
    journal = "Eur. Phys. J. C",
    volume = "80",
    number = "3",
    pages = "241",
    year = "2020",
    note = "[Erratum: Eur.Phys.J.C 80, 410 (2020)]"
}

@article{Gonzalez-Solis:2020jlh,
    author = "Gonz\`alez-Sol\'\i{}s, Sergi and Miranda, Alejandro and Rend\'on, Javier and Roig, Pablo",
    title = "{Exclusive hadronic tau decays as probes of non-SM interactions}",
    eprint = "1912.08725",
    archivePrefix = "arXiv",
    primaryClass = "hep-ph",
    doi = "10.1016/j.physletb.2020.135371",
    journal = "Phys. Lett. B",
    volume = "804",
    pages = "135371",
    year = "2020"
}

@article{Muong-2:2025xyk,
    author = "Aguillard, D. P. and others",
    collaboration = "Muon g-2",
    title = "{Measurement of the Positive Muon Anomalous Magnetic Moment to 127 ppb}",
    eprint = "2506.03069",
    archivePrefix = "arXiv",
    primaryClass = "hep-ex",
    reportNumber = "FERMILAB-PUB-25-0364-PPD",
    month = "6",
    year = "2025"
}

@article{Giusti:2019xct,
    author = "Giusti, D. and Lubicz, V. and Martinelli, G. and Sanfilippo, F. and Simula, S.",
    title = "{Electromagnetic and strong isospin-breaking corrections to the muon $g - 2$ from Lattice QCD+QED}",
    eprint = "1901.10462",
    archivePrefix = "arXiv",
    primaryClass = "hep-lat",
    doi = "10.1103/PhysRevD.99.114502",
    journal = "Phys. Rev. D",
    volume = "99",
    number = "11",
    pages = "114502",
    year = "2019"
}

@article{Borsanyi:2020mff,
    author = "Borsanyi, Sz. and others",
    title = "{Leading hadronic contribution to the muon magnetic moment from lattice QCD}",
    eprint = "2002.12347",
    archivePrefix = "arXiv",
    primaryClass = "hep-lat",
    doi = "10.1038/s41586-021-03418-1",
    journal = "Nature",
    volume = "593",
    number = "7857",
    pages = "51--55",
    year = "2021"
}

@article{Lehner:2020crt,
    author = "Lehner, Christoph and Meyer, Aaron S.",
    title = "{Consistency of hadronic vacuum polarization between lattice QCD and the R-ratio}",
    eprint = "2003.04177",
    archivePrefix = "arXiv",
    primaryClass = "hep-lat",
    doi = "10.1103/PhysRevD.101.074515",
    journal = "Phys. Rev. D",
    volume = "101",
    pages = "074515",
    year = "2020"
}

@article{Wang:2022lkq,
    author = "Wang, Gen and Draper, Terrence and Liu, Keh-Fei and Yang, Yi-Bo",
    collaboration = "chiQCD",
    title = "{Muon g-2 with overlap valence fermions}",
    eprint = "2204.01280",
    archivePrefix = "arXiv",
    primaryClass = "hep-lat",
    doi = "10.1103/PhysRevD.107.034513",
    journal = "Phys. Rev. D",
    volume = "107",
    number = "3",
    pages = "034513",
    year = "2023"
}

@article{Aubin:2022hgm,
    author = "Aubin, Christopher and Blum, Thomas and Golterman, Maarten and Peris, Santiago",
    title = "{Muon anomalous magnetic moment with staggered fermions: Is the lattice spacing small enough?}",
    eprint = "2204.12256",
    archivePrefix = "arXiv",
    primaryClass = "hep-lat",
    doi = "10.1103/PhysRevD.106.054503",
    journal = "Phys. Rev. D",
    volume = "106",
    number = "5",
    pages = "054503",
    year = "2022"
}

@article{Ce:2022kxy,
    author = "C\`e, Marco and others",
    title = "{Window observable for the hadronic vacuum polarization contribution to the muon g-2 from lattice QCD}",
    eprint = "2206.06582",
    archivePrefix = "arXiv",
    primaryClass = "hep-lat",
    reportNumber = "MITP-22-038, CERN-TH-2022-098, DESY-22-105",
    doi = "10.1103/PhysRevD.106.114502",
    journal = "Phys. Rev. D",
    volume = "106",
    number = "11",
    pages = "114502",
    year = "2022"
}

@article{ExtendedTwistedMass:2022jpw,
    author = "Alexandrou, C. and others",
    collaboration = "Extended Twisted Mass",
    title = "{Lattice calculation of the short and intermediate time-distance hadronic vacuum polarization contributions to the muon magnetic moment using twisted-mass fermions}",
    eprint = "2206.15084",
    archivePrefix = "arXiv",
    primaryClass = "hep-lat",
    doi = "10.1103/PhysRevD.107.074506",
    journal = "Phys. Rev. D",
    volume = "107",
    number = "7",
    pages = "074506",
    year = "2023"
}

@article{RBC:2023pvn,
    author = "Blum, T. and others",
    collaboration = "RBC, UKQCD",
    title = "{Update of Euclidean windows of the hadronic vacuum polarization}",
    eprint = "2301.08696",
    archivePrefix = "arXiv",
    primaryClass = "hep-lat",
    reportNumber = "CERN-TH-2023-010",
    doi = "10.1103/PhysRevD.108.054507",
    journal = "Phys. Rev. D",
    volume = "108",
    number = "5",
    pages = "054507",
    year = "2023"
}

@article{Kuberski:2024bcj,
    author = "Kuberski, Simon and C\`e, Marco and von Hippel, Georg and Meyer, Harvey B. and Ottnad, Konstantin and Risch, Andreas and Wittig, Hartmut",
    title = "{Hadronic vacuum polarization in the muon g \ensuremath{-} 2: the short-distance contribution from lattice QCD}",
    eprint = "2401.11895",
    archivePrefix = "arXiv",
    primaryClass = "hep-lat",
    reportNumber = "MITP-24-011, CERN-TH-2024-011",
    doi = "10.1007/JHEP03(2024)172",
    journal = "JHEP",
    volume = "03",
    pages = "172",
    year = "2024"
}

@article{Boccaletti:2024guq,
    author = "Boccaletti, A. and others",
    title = "{High precision calculation of the hadronic vacuum polarisation contribution to the muon anomaly}",
    eprint = "2407.10913",
    archivePrefix = "arXiv",
    primaryClass = "hep-lat",
    month = "7",
    year = "2024"
}

@article{Spiegel:2024dec,
    author = "Spiegel, Sebastian and Lehner, Christoph",
    title = "{A high-precision continuum limit study of the HVP short-distance window}",
    eprint = "2410.17053",
    archivePrefix = "arXiv",
    primaryClass = "hep-lat",
    month = "10",
    year = "2024"
}

@article{RBC:2024fic,
    author = "Blum, T. and others",
    collaboration = "RBC, UKQCD",
    title = "{Long-Distance Window of the Hadronic Vacuum Polarization for the Muon g-2}",
    eprint = "2410.20590",
    archivePrefix = "arXiv",
    primaryClass = "hep-lat",
    reportNumber = "CERN-TH-2024-182",
    doi = "10.1103/PhysRevLett.134.201901",
    journal = "Phys. Rev. Lett.",
    volume = "134",
    number = "20",
    pages = "201901",
    year = "2025"
}

@article{Djukanovic:2024cmq,
    author = "Djukanovic, Dalibor and von Hippel, Georg and Kuberski, Simon and Meyer, Harvey B. and Miller, Nolan and Ottnad, Konstantin and Parrino, Julian and Risch, Andreas and Wittig, Hartmut",
    title = "{The hadronic vacuum polarization contribution to the muon g \ensuremath{-} 2 at long distances}",
    eprint = "2411.07969",
    archivePrefix = "arXiv",
    primaryClass = "hep-lat",
    reportNumber = "CERN-TH-2024-196, MITP-24-080",
    doi = "10.1007/JHEP04(2025)098",
    journal = "JHEP",
    volume = "04",
    pages = "098",
    year = "2025"
}

@article{ExtendedTwistedMass:2024nyi,
    author = "Alexandrou, C. and others",
    collaboration = "Extended Twisted Mass",
    title = "{Strange and charm quark contributions to the muon anomalous magnetic moment in lattice QCD with twisted-mass fermions}",
    eprint = "2411.08852",
    archivePrefix = "arXiv",
    primaryClass = "hep-lat",
    reportNumber = "CERN-TH-2024-197",
    doi = "10.1103/PhysRevD.111.054502",
    journal = "Phys. Rev. D",
    volume = "111",
    number = "5",
    pages = "054502",
    year = "2025"
}

@article{MILC:2024ryz,
    author = "Bazavov, Alexei and others",
    collaboration = "MILC, Fermilab Lattice, HPQCD",
    title = "{Hadronic vacuum polarization for the muon g-2 from lattice QCD: Complete short and intermediate windows}",
    eprint = "2411.09656",
    archivePrefix = "arXiv",
    primaryClass = "hep-lat",
    reportNumber = "FERMILAB-PUB-24-0835-T",
    doi = "10.1103/PhysRevD.111.094508",
    journal = "Phys. Rev. D",
    volume = "111",
    number = "9",
    pages = "094508",
    year = "2025"
}

@article{Bazavov:2024eou,
    author = "Bazavov, Alexei and others",
    title = "{Hadronic vacuum polarization for the muon $g-2$ from lattice QCD: Long-distance and full light-quark connected contribution}",
    eprint = "2412.18491",
    archivePrefix = "arXiv",
    primaryClass = "hep-lat",
    reportNumber = "FERMILAB-PUB-24-0957-T",
    month = "12",
    year = "2024"
}

@article{Keshavarzi:2019abf,
    author = "Keshavarzi, Alexander and Nomura, Daisuke and Teubner, Thomas",
    title = "{$g-2$ of charged leptons, $\alpha (M^2_Z)$ , and the hyperfine splitting of muonium}",
    eprint = "1911.00367",
    archivePrefix = "arXiv",
    primaryClass = "hep-ph",
    reportNumber = "MAN/HEP/2019/010, LTH 1216, KEK-TH-2165",
    doi = "10.1103/PhysRevD.101.014029",
    journal = "Phys. Rev. D",
    volume = "101",
    number = "1",
    pages = "014029",
    year = "2020"
}

@article{DiLuzio:2024sps,
    author = "Di Luzio, Luca and Keshavarzi, Alexander and Masiero, Antonio and Paradisi, Paride",
    title = "{Model-Independent Tests of the Hadronic Vacuum Polarization Contribution to the Muon g-2}",
    eprint = "2408.01123",
    archivePrefix = "arXiv",
    primaryClass = "hep-ph",
    doi = "10.1103/PhysRevLett.134.011902",
    journal = "Phys. Rev. Lett.",
    volume = "134",
    number = "1",
    pages = "011902",
    year = "2025"
}

@article{Kurz:2014wya,
    author = "Kurz, Alexander and Liu, Tao and Marquard, Peter and Steinhauser, Matthias",
    title = "{Hadronic contribution to the muon anomalous magnetic moment to next-to-next-to-leading order}",
    eprint = "1403.6400",
    archivePrefix = "arXiv",
    primaryClass = "hep-ph",
    reportNumber = "SFB-CPP-14-19, TTP14-009, DESY-14-038, LPN14-056",
    doi = "10.1016/j.physletb.2014.05.043",
    journal = "Phys. Lett. B",
    volume = "734",
    pages = "144--147",
    year = "2014"
}

@article{Colangelo:2015ama,
    author = "Colangelo, Gilberto and Hoferichter, Martin and Procura, Massimiliano and Stoffer, Peter",
    title = "{Dispersion relation for hadronic light-by-light scattering: theoretical foundations}",
    eprint = "1506.01386",
    archivePrefix = "arXiv",
    primaryClass = "hep-ph",
    reportNumber = "INT-PUB-15-019, UWTHPH-2015-10",
    doi = "10.1007/JHEP09(2015)074",
    journal = "JHEP",
    volume = "09",
    pages = "074",
    year = "2015"
}

@article{Bijnens:2019ghy,
    author = "Bijnens, Johan and Hermansson-Truedsson, Nils and Rodr\'\i{}guez-S\'anchez, Antonio",
    title = "{Short-distance constraints for the HLbL contribution to the muon anomalous magnetic moment}",
    eprint = "1908.03331",
    archivePrefix = "arXiv",
    primaryClass = "hep-ph",
    reportNumber = "LU TP 19-38",
    doi = "10.1016/j.physletb.2019.134994",
    journal = "Phys. Lett. B",
    volume = "798",
    pages = "134994",
    year = "2019"
}

@article{Bijnens:2020xnl,
    author = "Bijnens, Johan and Hermansson-Truedsson, Nils and Laub, Laetitia and Rodr\'\i{}guez-S\'anchez, Antonio",
    title = "{Short-distance HLbL contributions to the muon anomalous magnetic moment beyond perturbation theory}",
    eprint = "2008.13487",
    archivePrefix = "arXiv",
    primaryClass = "hep-ph",
    reportNumber = "LU TP 20-47",
    doi = "10.1007/JHEP10(2020)203",
    journal = "JHEP",
    volume = "10",
    pages = "203",
    year = "2020"
}

@article{Bijnens:2021jqo,
    author = "Bijnens, Johan and Hermansson-Truedsson, Nils and Laub, Laetitia and Rodr\'\i{}guez-S\'anchez, Antonio",
    title = "{The two-loop perturbative correction to the $(g - 2)_\mu$ HLbL at short distances}",
    eprint = "2101.09169",
    archivePrefix = "arXiv",
    primaryClass = "hep-ph",
    reportNumber = "LU TP 21-03",
    doi = "10.1007/JHEP04(2021)240",
    journal = "JHEP",
    volume = "04",
    pages = "240",
    year = "2021"
}

@article{Leutgeb:2022lqw,
    author = "Leutgeb, Josef and Mager, Jonas and Rebhan, Anton",
    title = "{Hadronic light-by-light contribution to the muon g-2 from holographic QCD with solved U(1)A problem}",
    eprint = "2211.16562",
    archivePrefix = "arXiv",
    primaryClass = "hep-ph",
    doi = "10.1103/PhysRevD.107.054021",
    journal = "Phys. Rev. D",
    volume = "107",
    number = "5",
    pages = "054021",
    year = "2023"
}

@article{Hoferichter:2023tgp,
    author = "Hoferichter, Martin and Kubis, Bastian and Zanke, Marvin",
    title = "{Axial-vector transition form factors and $e^+ e^- \to f_1 \pi^+ \pi^-$}",
    eprint = "2307.14413",
    archivePrefix = "arXiv",
    primaryClass = "hep-ph",
    doi = "10.1007/JHEP08(2023)209",
    journal = "JHEP",
    volume = "08",
    pages = "209",
    year = "2023"
}

@article{Hoferichter:2024fsj,
    author = "Hoferichter, Martin and Stoffer, Peter and Zillinger, Maximilian",
    title = "{An optimized basis for hadronic light-by-light scattering}",
    eprint = "2402.14060",
    archivePrefix = "arXiv",
    primaryClass = "hep-ph",
    reportNumber = "PSI-PR-24-08, ZU-TH 11/24",
    doi = "10.1007/JHEP04(2024)092",
    journal = "JHEP",
    volume = "04",
    pages = "092",
    year = "2024"
}

@article{Ludtke:2024ase,
    author = {L\"udtke, Jan and Procura, Massimiliano and Stoffer, Peter},
    title = "{Dispersion relations for the hadronic VVA correlator}",
    eprint = "2410.11946",
    archivePrefix = "arXiv",
    primaryClass = "hep-ph",
    reportNumber = "PSI-PR-24-18, UWThPh 2024-20, ZU-TH 49/24",
    doi = "10.1007/JHEP04(2025)130",
    journal = "JHEP",
    volume = "04",
    pages = "130",
    year = "2025"
}

@article{Deineka:2024mzt,
    author = "Deineka, Oleksandra and Danilkin, Igor and Vanderhaeghen, Marc",
    title = "{Dispersive estimate of the a0(980) contribution to (g-2)\ensuremath{\mu}}",
    eprint = "2410.12894",
    archivePrefix = "arXiv",
    primaryClass = "hep-ph",
    doi = "10.1103/PhysRevD.111.034009",
    journal = "Phys. Rev. D",
    volume = "111",
    number = "3",
    pages = "034009",
    year = "2025"
}

@article{Bijnens:2024jgh,
    author = "Bijnens, Johan and Hermansson-Truedsson, Nils and Rodr\'\i{}guez-S\'anchez, Antonio",
    title = "{Constraints on the hadronic light-by-light tensor in corner kinematics for the muon g \ensuremath{-} 2}",
    eprint = "2411.09578",
    archivePrefix = "arXiv",
    primaryClass = "hep-ph",
    doi = "10.1007/JHEP03(2025)094",
    journal = "JHEP",
    volume = "03",
    pages = "094",
    year = "2025"
}

@article{Colangelo:2014qya,
    author = "Colangelo, Gilberto and Hoferichter, Martin and Nyffeler, Andreas and Passera, Massimo and Stoffer, Peter",
    title = "{Remarks on higher-order hadronic corrections to the muon g\ensuremath{-}2}",
    eprint = "1403.7512",
    archivePrefix = "arXiv",
    primaryClass = "hep-ph",
    doi = "10.1016/j.physletb.2014.06.012",
    journal = "Phys. Lett. B",
    volume = "735",
    pages = "90--91",
    year = "2014"
}

@article{Blum:2019ugy,
    author = "Blum, Thomas and Christ, Norman and Hayakawa, Masashi and Izubuchi, Taku and Jin, Luchang and Jung, Chulwoo and Lehner, Christoph",
    title = "{Hadronic Light-by-Light Scattering Contribution to the Muon Anomalous Magnetic Moment from Lattice QCD}",
    eprint = "1911.08123",
    archivePrefix = "arXiv",
    primaryClass = "hep-lat",
    doi = "10.1103/PhysRevLett.124.132002",
    journal = "Phys. Rev. Lett.",
    volume = "124",
    number = "13",
    pages = "132002",
    year = "2020"
}

@article{Chao:2021tvp,
    author = "Chao, En-Hung and Hudspith, Renwick J. and G\'erardin, Antoine and Green, Jeremy R. and Meyer, Harvey B. and Ottnad, Konstantin",
    title = "{Hadronic light-by-light contribution to $(g-2)_\mu $ from lattice QCD: a complete calculation}",
    eprint = "2104.02632",
    archivePrefix = "arXiv",
    primaryClass = "hep-lat",
    doi = "10.1140/epjc/s10052-021-09455-4",
    journal = "Eur. Phys. J. C",
    volume = "81",
    number = "7",
    pages = "651",
    year = "2021"
}

@article{Chao:2022xzg,
    author = "Chao, En-Hung and Hudspith, Renwick J. and G\'erardin, Antoine and Green, Jeremy R. and Meyer, Harvey B.",
    title = "{The charm-quark contribution to light-by-light scattering in the muon $(g-2)$ from lattice QCD}",
    eprint = "2204.08844",
    archivePrefix = "arXiv",
    primaryClass = "hep-lat",
    reportNumber = "MITP-22-031",
    doi = "10.1140/epjc/s10052-022-10589-2",
    journal = "Eur. Phys. J. C",
    volume = "82",
    number = "8",
    pages = "664",
    year = "2022"
}

@article{Blum:2023vlm,
    author = "Blum, Thomas and Christ, Norman and Hayakawa, Masashi and Izubuchi, Taku and Jin, Luchang and Jung, Chulwoo and Lehner, Christoph and Tu, Cheng",
    collaboration = "RBC, UKQCD",
    title = "{Hadronic light-by-light contribution to the muon anomaly from lattice QCD with infinite volume QED at physical pion mass}",
    eprint = "2304.04423",
    archivePrefix = "arXiv",
    primaryClass = "hep-lat",
    doi = "10.1103/PhysRevD.111.014501",
    journal = "Phys. Rev. D",
    volume = "111",
    number = "1",
    pages = "014501",
    year = "2025"
}

@article{Fodor:2024jyn,
    author = "Fodor, Zoltan and Gerardin, Antoine and Lellouch, Laurent and Szabo, Kalman K. and Toth, Balint C. and Zimmermann, Christian",
    title = "{Hadronic light-by-light scattering contribution to the anomalous magnetic moment of the muon at the physical pion mass}",
    eprint = "2411.11719",
    archivePrefix = "arXiv",
    primaryClass = "hep-lat",
    month = "11",
    year = "2024"
}

@article{Aoyama:2012wk,
    author = "Aoyama, Tatsumi and Hayakawa, Masashi and Kinoshita, Toichiro and Nio, Makiko",
    title = "{Complete Tenth-Order QED Contribution to the Muon g-2}",
    eprint = "1205.5370",
    archivePrefix = "arXiv",
    primaryClass = "hep-ph",
    reportNumber = "RIKEN-QHP-26",
    doi = "10.1103/PhysRevLett.109.111808",
    journal = "Phys. Rev. Lett.",
    volume = "109",
    pages = "111808",
    year = "2012"
}

@article{Volkov:2019phy,
    author = "Volkov, Sergey",
    title = "{Calculating the five-loop QED contribution to the electron anomalous magnetic moment: Graphs without lepton loops}",
    eprint = "1909.08015",
    archivePrefix = "arXiv",
    primaryClass = "hep-ph",
    doi = "10.1103/PhysRevD.100.096004",
    journal = "Phys. Rev. D",
    volume = "100",
    number = "9",
    pages = "096004",
    year = "2019"
}

@article{Volkov:2024yzc,
    author = "Volkov, Sergey",
    title = "{Calculation of the total 10th order QED contribution to the electron magnetic moment}",
    eprint = "2404.00649",
    archivePrefix = "arXiv",
    primaryClass = "hep-ph",
    doi = "10.1103/PhysRevD.110.036001",
    journal = "Phys. Rev. D",
    volume = "110",
    number = "3",
    pages = "036001",
    year = "2024"
}

@article{Aoyama:2024aly,
    author = "Aoyama, Tatsumi and Hayakawa, Masashi and Hirayama, Akira and Nio, Makiko",
    title = "{Verification of the tenth-order QED contribution to the anomalous magnetic moment of the electron from diagrams without fermion loops}",
    eprint = "2412.06473",
    archivePrefix = "arXiv",
    primaryClass = "hep-ph",
    doi = "10.1103/PhysRevD.111.L031902",
    journal = "Phys. Rev. D",
    volume = "111",
    number = "3",
    pages = "L031902",
    year = "2025"
}

@article{Parker:2018vye,
    author = {Parker, Richard H. and Yu, Chenghui and Zhong, Weicheng and Estey, Brian and M\"uller, Holger},
    title = "{Measurement of the fine-structure constant as a test of the Standard Model}",
    eprint = "1812.04130",
    archivePrefix = "arXiv",
    primaryClass = "physics.atom-ph",
    doi = "10.1126/science.aap7706",
    journal = "Science",
    volume = "360",
    pages = "191",
    year = "2018"
}

@article{Morel:2020dww,
    author = {Morel, L\'eo and Yao, Zhibin and Clad\'e, Pierre and Guellati-Kh\'elifa, Sa\"\i{}da},
    title = "{Determination of the fine-structure constant with an accuracy of 81 parts per trillion}",
    doi = "10.1038/s41586-020-2964-7",
    journal = "Nature",
    volume = "588",
    number = "7836",
    pages = "61--65",
    year = "2020"
}

@article{Fan:2022eto,
    author = "Fan, X. and Myers, T. G. and Sukra, B. A. D. and Gabrielse, G.",
    title = "{Measurement of the Electron Magnetic Moment}",
    eprint = "2209.13084",
    archivePrefix = "arXiv",
    primaryClass = "physics.atom-ph",
    doi = "10.1103/PhysRevLett.130.071801",
    journal = "Phys. Rev. Lett.",
    volume = "130",
    number = "7",
    pages = "071801",
    year = "2023"
}

@article{Czarnecki:2002nt,
    author = "Czarnecki, Andrzej and Marciano, William J. and Vainshtein, Arkady",
    title = "{Refinements in electroweak contributions to the muon anomalous magnetic moment}",
    eprint = "hep-ph/0212229",
    archivePrefix = "arXiv",
    reportNumber = "ALBERTA-THY-18-02, BNL-HET-02-25, TPI-MINN-02-46, UMN-TH-2119-02",
    doi = "10.1103/PhysRevD.67.073006",
    journal = "Phys. Rev. D",
    volume = "67",
    pages = "073006",
    year = "2003",
    note = "[Erratum: Phys.Rev.D 73, 119901 (2006)]"
}

@article{Gnendiger:2013pva,
    author = {Gnendiger, C. and St\"ockinger, D. and St\"ockinger-Kim, H.},
    title = "{The electroweak contributions to $(g-2)_\mu$ after the Higgs boson mass measurement}",
    eprint = "1306.5546",
    archivePrefix = "arXiv",
    primaryClass = "hep-ph",
    doi = "10.1103/PhysRevD.88.053005",
    journal = "Phys. Rev. D",
    volume = "88",
    pages = "053005",
    year = "2013"
}

@article{Hoferichter:2025yih,
    author = {Hoferichter, Martin and L\"udtke, Jan and Naterop, Luca and Procura, Massimiliano and Stoffer, Peter},
    title = "{Improved Evaluation of the Electroweak Contribution to Muon g-2}",
    eprint = "2503.04883",
    archivePrefix = "arXiv",
    primaryClass = "hep-ph",
    reportNumber = "PSI-PR-25-04, UWThPh 2025-8, ZU-TH 10/25",
    doi = "10.1103/PhysRevLett.134.201801",
    journal = "Phys. Rev. Lett.",
    volume = "134",
    number = "20",
    pages = "201801",
    year = "2025"
}

@article{Masjuan:2017tvw,
    author = "Masjuan, Pere and Sánchez-Puertas, Pablo",
    title = "{Pseudoscalar-pole contribution to the $(g_{\mu}-2)$: a rational approach}",
    eprint = "1701.05829",
    archivePrefix = "arXiv",
    primaryClass = "hep-ph",
    doi = "10.1103/PhysRevD.95.054026",
    journal = "Phys. Rev. D",
    volume = "95",
    number = "5",
    pages = "054026",
    year = "2017"
}

@article{Colangelo:2017fiz,
    author = "Colangelo, Gilberto and Hoferichter, Martin and Procura, Massimiliano and Stoffer, Peter",
    title = "{Dispersion relation for hadronic light-by-light scattering: two-pion contributions}",
    eprint = "1702.07347",
    archivePrefix = "arXiv",
    primaryClass = "hep-ph",
    reportNumber = "INT-PUB-17-009, CERN-TH-2017-041, NSF-KITP-17-036",
    doi = "10.1007/JHEP04(2017)161",
    journal = "JHEP",
    volume = "04",
    pages = "161",
    year = "2017"
}

@article{Eichmann:2019tjk,
    author = "Eichmann, Gernot and Fischer, Christian S. and Weil, Esther and Williams, Richard",
    title = "{Single pseudoscalar meson pole and pion box contributions to the anomalous magnetic moment of the muon}",
    eprint = "1903.10844",
    archivePrefix = "arXiv",
    primaryClass = "hep-ph",
    doi = "10.1016/j.physletb.2019.134855",
    journal = "Phys. Lett. B",
    volume = "797",
    pages = "134855",
    year = "2019",
    note = "[Erratum: Phys.Lett.B 799, 135029 (2019)]"
}

@article{Cappiello:2019hwh,
    author = "Cappiello, Luigi and Cat\`a, Oscar and D'Ambrosio, Giancarlo and Greynat, David and Iyer, Abhishek",
    title = "{Axial-vector and pseudoscalar mesons in the hadronic light-by-light contribution to the muon $(g-2)$}",
    eprint = "1912.02779",
    archivePrefix = "arXiv",
    primaryClass = "hep-ph",
    doi = "10.1103/PhysRevD.102.016009",
    journal = "Phys. Rev. D",
    volume = "102",
    number = "1",
    pages = "016009",
    year = "2020"
}

@article{Estrada:2024cfy,
    author = "Estrada, Emilio J. and Gonz\`alez-Sol\'\i{}s, Sergi and Guevara, Adolfo and Roig, Pablo",
    title = "{Improved \ensuremath{\pi}$^{0}$, \ensuremath{\eta}, \ensuremath{\eta}' transition form factors in resonance chiral theory and their $ {a}_{\mu}^{\textrm{HLbL}} $ contribution}",
    eprint = "2409.10503",
    archivePrefix = "arXiv",
    primaryClass = "hep-ph",
    doi = "10.1007/JHEP12(2024)203",
    journal = "JHEP",
    volume = "12",
    pages = "203",
    year = "2024"
}

@article{CERN-Mainz-Daresbury:1978ccd,
    author = "Bailey, J. and others",
    collaboration = "CERN-Mainz-Daresbury",
    title = "{Final Report on the CERN Muon Storage Ring Including the Anomalous Magnetic Moment and the Electric Dipole Moment of the Muon, and a Direct Test of Relativistic Time Dilation}",
    reportNumber = "Print-79-0088 (CERN)",
    doi = "10.1016/0550-3213(79)90292-X",
    journal = "Nucl. Phys. B",
    volume = "150",
    pages = "1--75",
    year = "1979"
}

@article{Aoyama:2020ynm,
    author = "Aoyama, T. and others",
    title = "{The anomalous magnetic moment of the muon in the Standard Model}",
    eprint = "2006.04822",
    archivePrefix = "arXiv",
    primaryClass = "hep-ph",
    reportNumber = "FERMILAB-PUB-20-207-T, INT-PUB-20-021, KEK Preprint 2020-5,
  MITP/20-028, KEK Preprint 2020-5, MITP/20-028, CERN-TH-2020-075, IFT-UAM/CSIC-20-74, LMU-ASC 18/20, LTH 1234,
  LU TP 20-20, LTH 1234, LU TP 20-20, MAN/HEP/2020/003, PSI-PR-20-06, UWThPh 2020-14, ZU-TH 18/20",
    doi = "10.1016/j.physrep.2020.07.006",
    journal = "Phys. Rept.",
    volume = "887",
    pages = "1--166",
    year = "2020"
}

@article{Fritzsch:1973pi,
    author = "Fritzsch, H. and Gell-Mann, Murray and Leutwyler, H.",
    title = "{Advantages of the Color Octet Gluon Picture}",
    reportNumber = "CALT-68-409",
    doi = "10.1016/0370-2693(73)90625-4",
    journal = "Phys. Lett. B",
    volume = "47",
    pages = "365--368",
    year = "1973"
}

@article{Gross:1973id,
    author = "Gross, David J. and Wilczek, Frank",
    editor = "Taylor, J. C.",
    title = "{Ultraviolet Behavior of Nonabelian Gauge Theories}",
    doi = "10.1103/PhysRevLett.30.1343",
    journal = "Phys. Rev. Lett.",
    volume = "30",
    pages = "1343--1346",
    year = "1973"
}

@article{Politzer:1973fx,
    author = "Politzer, H. David",
    editor = "Taylor, J. C.",
    title = "{Reliable Perturbative Results for Strong Interactions?}",
    doi = "10.1103/PhysRevLett.30.1346",
    journal = "Phys. Rev. Lett.",
    volume = "30",
    pages = "1346--1349",
    year = "1973"
}

@article{Wilson:1974sk,
    author = "Wilson, Kenneth G.",
    editor = "Taylor, J. C.",
    title = "{Confinement of Quarks}",
    reportNumber = "CLNS-262",
    doi = "10.1103/PhysRevD.10.2445",
    journal = "Phys. Rev. D",
    volume = "10",
    pages = "2445--2459",
    year = "1974"
}

@article{Weinberg:1978kz,
    author = "Weinberg, Steven",
    editor = "Deser, S.",
    title = "{Phenomenological Lagrangians}",
    reportNumber = "HUTP-78-A051A",
    doi = "10.1016/0378-4371(79)90223-1",
    journal = "Physica A",
    volume = "96",
    number = "1-2",
    pages = "327--340",
    year = "1979"
}

@article{Gasser:1984gg,
    author = "Gasser, J. and Leutwyler, H.",
    title = "{Chiral Perturbation Theory: Expansions in the Mass of the Strange Quark}",
    reportNumber = "CERN-TH-3798",
    doi = "10.1016/0550-3213(85)90492-4",
    journal = "Nucl. Phys. B",
    volume = "250",
    pages = "465--516",
    year = "1985"
}

@article{Gasser:1983yg,
    author = "Gasser, J. and Leutwyler, H.",
    title = "{Chiral Perturbation Theory to One Loop}",
    reportNumber = "CERN-TH-3689",
    doi = "10.1016/0003-4916(84)90242-2",
    journal = "Annals Phys.",
    volume = "158",
    pages = "142",
    year = "1984"
}

@article{Bijnens:1999sh,
    author = "Bijnens, Johan and Colangelo, Gilberto and Ecker, Gerhard",
    title = "{The Mesonic chiral Lagrangian of order p**6}",
    eprint = "hep-ph/9902437",
    archivePrefix = "arXiv",
    reportNumber = "LU-TP-99-02, UWTHPH-1999-02, ZU-TH-9-99",
    doi = "10.1088/1126-6708/1999/02/020",
    journal = "JHEP",
    volume = "02",
    pages = "020",
    year = "1999"
}

@article{Bijnens:2001bb,
    author = "Bijnens, J. and Girlanda, L. and Talavera, P.",
    title = "{The Anomalous chiral Lagrangian of order p**6}",
    eprint = "hep-ph/0110400",
    archivePrefix = "arXiv",
    reportNumber = "LU-TP-01-34, DFPD-01-TH-25, CPT-2001-P4256",
    doi = "10.1007/s100520100887",
    journal = "Eur. Phys. J. C",
    volume = "23",
    pages = "539--544",
    year = "2002"
}

\end{document}